\documentclass{emulateapj}

\usepackage{apjfonts}

\newcommand{\tnm}{\tablenotemark}
\newcommand{\tnt}{\tablenotetext}

\newcommand{\ergs}{erg s$^{-1}$}
\newcommand{\flux}{erg cm$^{-2}$ s$^{-1}$}

\newcommand{\xmmnewton}{{\it XMM-Newton}}
\newcommand{\suzaku}{{\it Suzaku}}
\newcommand{\chandra}{{\it Chandra}}

\newcommand{\rxte}{{\it RXTE}}

\newcommand{\beq}{\begin{equation}}
\newcommand{\eeq}{\end{equation}}

\newcommand{\alfven}{Alfv\'{e}n}

\newcommand{\dgr}{$^{\circ}$}

\newcommand{\msun}{M_{\sun}}

\usepackage{psfig}

\begin{document}
\slugcomment{Accepted for publication in The Astrophysical Journal}


\title{{\it SUZAKU} X-RAY SPECTRA AND PULSE PROFILE VARIATIONS DURING THE SUPERORBITAL CYCLE OF LMC X-4}
\shorttitle{{\it SUZAKU} X-RAY SPECTRA AND PULSE PROFILES OF LMC X-4}
\shortauthors{HUNG, HICKOX, BOROSON, \& VRTILEK}
\author{Li-Wei Hung\altaffilmark{1,2}, Ryan C. Hickox\altaffilmark{2,3}, Bram S. Boroson\altaffilmark{2}, Saeqa D. Vrtilek\altaffilmark{2}}

\altaffiltext{1}{Department of Astronomy, The Ohio State University, McPherson Laboratory 140 W 18th Avenue, Columbus, Ohio 43210-1173, USA; hung.88@buckeyemail.osu.edu}

\altaffiltext{2}{Harvard-Smithsonian Center for Astrophysics, 60 Garden Street,
 Cambridge, MA 02138, USA; rhickox@cfa.harvard.edu,
 bboroson@cfa.harvard.edu, svrtilek@cfa.harvard.edu.}

\altaffiltext{3}{Department of Physics, Durham University, South Road, Durham, DH1 3LE, UK}

\begin{abstract}

  We present results from spectral and temporal analyses of
  \suzaku\ and \rxte\ observations of the high-mass X-ray binary LMC
  X-4.  Using the full 13 years of available \rxte/all-sky monitor data, we apply
  the ANOVA and Lomb Normalized Periodogram methods to obtain an
  improved superorbital period measurement of $30.32\pm0.04$ days.
  The phase-averaged X-ray spectra from \suzaku\ observations during
  the high state of the superorbital period can be modeled in the
  0.6--50 keV band as the combination of a power law with $\Gamma \sim
  0.6$ and a high-energy cutoff at $\sim$25 keV, a blackbody with
  $kT_{\rm BB} \sim 0.18\ \rm keV$, and emission lines from Fe
  K$_\alpha$, O\,{\sc viii}, and Ne\,{\sc ix} ({\sc x} Ly$\alpha$).  Assuming a distance
  of 50 kpc, the source has luminosity $L_{\rm X}\sim 3 \times
  10^{38}$ \ergs\ in the 2--50 keV band, and the luminosity of the
  soft (blackbody) component is $L_{\rm BB}\sim 1.5\times 10^{37}$
  \ergs.  The energy-resolved pulse profiles show single-peaked soft
  (0.5--1 keV) and hard (6--10 keV) pulses but a more complex pattern
  of medium (2--10 keV) pulses; cross-correlation of the hard with the
  soft pulses shows a phase shift that varies between observations.
  We interpret these results in terms of a picture in which a
  precessing disk reprocesses the hard X-rays and produces the
  observed soft spectral component, as has been suggested for the
  similar sources Her X-1 and SMC X-1.

\end{abstract}


\section{Introduction}

An X-ray binary system consists of a normal star orbiting a compact
object (neutron star or black hole) to which the normal star
transfers matter either by overflowing its Roche lobe or via a stellar
wind.  X-ray pulsars are a class of X-ray binaries in which the
neutron star has a strong magnetic field, so that accreting matter
follows the field lines and falls into the magnetic poles, generating
pulses as the magnetic poles rotate in and out of our line of sight.
While the general picture of the accretion mechanism is well known,
the physics of accretion near the magnetosphere, where the neutron
star's magnetic field begins to dominate the flow, is not fully
understood.  By studying individual X-ray pulsars in detail, we hope
to gain a better understanding of the accretion mechanism.

In this paper, we study the X-ray pulsar LMC X-4, which is a high-mass
X-ray binary system consisting of a 1.25 $\msun$ neutron star
accreting from a 14.5 $\msun$ O8 III companion \citep{kell83,meer07}.
The neutron star rotates with a period of $\sim$13.5\ s, and orbits
its companion with a period of $\sim$1.4\ days \citep{whit78}.  In
addition to the pulse period and the orbital period, the system has
been observed to have a long-term period of $30.5\pm0.5$ days
\citep{lang81}.  Between the high and low states of this long-term
``superorbital'' period (believed to be caused by a precessing
accretion disk that periodically obscures the neutron star), the X-ray
flux varies by 2 orders of magnitude \citep{heem89,naik03}.  The
spectrum of LMC X-4 in the 0.1--100 keV band is well described by a
hard power law with a high-energy cutoff, a soft X-ray excess, and a
strong iron emission line \citep{naik03,laba01}.

\begin{figure}
\plotone{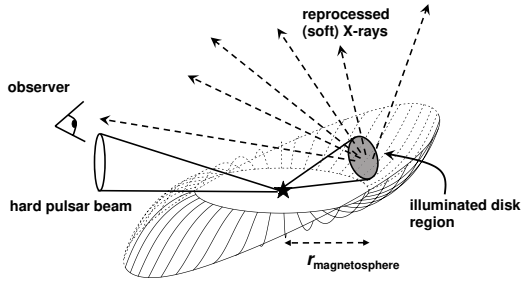}
\caption{Schematic of disk reprocessing in X-ray pulsars
  \citep{hick05}.  The hard (power-law spectrum) X-ray beam sweeps
  around and illuminates the accretion disk, which re-radiates a soft
  X-ray (blackbody) component.  The observer sees pulses from both
  power-law and blackbody components, but these differ in pulse shape
  and phase.  \label{sketch}}
\end{figure}

\citet{hick04} showed that a soft X-ray excess was a common, and
possibly ubiquitous, feature of X-ray pulsars, and that in
high-luminosity systems (with $L_X \gtrsim 10^{36}$ \ergs), the soft
excess originates from reprocessing of hard X-rays from the neutron
star by the inner accretion disk (Figure~\ref{sketch}).  This
interpretation had been put forward previously to explain the emission
from several X-ray pulsars; for example, \citet{burd00} suggested that
in Cen X-3, a soft excess below 1 keV originates from reprocessing of
the primary radiation by an opaque shell located at the magnetosphere.  In this
picture, the soft X-rays may be expected to show pulsations that
differ in shape and phase from the hard X-ray pulses.  \citet{endo00}
showed that the precessing accretion disk in the X-ray pulsar Her X-1
reprocessed hard X-rays ($> 2$ keV) from the pulsar and emitted soft
X-rays ($< 1$ keV), and \citet{zane04} showed that the phase
difference between hard and soft pulses changed as the accretion disk
precesses.  For the X-ray pulsar SMC X-1, a high-mass X-ray binary
similar to LMC X-4, \citet{neil04} suggested an explanation for the
observed characteristics of the soft pulses in terms of precession of
the accretion disk.  \citet{hick05} used a model of a twisted inner
disk, illuminated by the rotating X-ray pulsar beam, to simulate
pulsations in the soft component due to this reprocessing.  They found
that for some disk and beam geometries, precession of an illuminated
accretion disk can roughly reproduce the soft pulse profiles.
However, modeling the disk geometry of SMC X-1 is difficult since its
superorbital period changes with time.  The superorbital period of LMC
X-4 is more stable and hence may allow a better test of this picture.

In this paper, we study the X-ray spectra and pulse properties of LMC
X-4 at three different phases in the high state of its superorbital
cycle.  First, we determine an improved value for the superorbital
period based on 13 years of {\em Rossi X-Ray Timing Explorer} (\rxte)
all-sky monitor (ASM) data and use it to accurately determine the
superorbital phase of our three \suzaku\ observations.  We then
analyze the phase-averaged X-ray spectra and energy-resolved pulse
profiles for the \suzaku\ observations and interpret them in terms of
a simple model based on the reprocessing of hard X-rays by the
precessing accretion disk.  In Section~\ref{obs}, we describe the
observations, and in Section~\ref{analysis} we present our analysis and
results.  In Section~\ref{discussion}, we discuss the implications of our
results and future work.  All quoted uncertainties are 90\% confidence
for a single interesting parameter.


\section{Observations and Data Reduction}
\label{obs}

\subsection{{\it Suzaku}}
\label{subsec:suzakuobs}

\suzaku\ is an X-ray observatory launched in 2005 \citep{mits07},
covering the 0.2--600 keV energy range with two sets of instruments:
X-ray CCDs (X-ray Imaging Spectrometer: XIS) covering 0.2--12 keV and
the Hard X-ray Detector (HXD), a photon-counting (non-imaging)
detector covering 10--600 keV.  The HXD is comprised of two
instruments, the PIN (sensitive to energies 10--70 keV) and GSO
(sensitive to 40--600 keV).  Here, our analysis focuses on XIS
observations in the 0.5--10 keV band and the PIN in the 10--50 keV
band.  We do not make use of GSO data as the signal in the GSO is very
weak ($\lesssim1$\% of background).  Our analysis uses XIS and PIN
event files that have been reprocessed with the FTOOL {\tt aepipeline}
using the latest {\em Suzaku} calibration (version 2.2.7.18) and
screened using the standard criteria.

\suzaku\ performed three observations of LMC X-4 during the high state
of the superorbital cycle, when the source is bright enough to obtain
a high signal-to-noise ratio.  Details of the observations are shown
in Table\ \ref{suzakuobs}.  These observations were carried out at the
``XIS nominal'' pointing position for effective exposures of $\sim$20
ks.  Table \ref{suzakuobs} also gives the approximate orbital phases
covered by the observations, estimated from the ephemeris of
\citet{levi00}.  The exposures all avoid orbital phases close to
eclipse (the range of $\phi_{\rm orb}$ is 0.1--0.8) and do not
show any significant variations in flux other than the observed
pulsations, so we do not expect any changes in absorption (from the
companion star or accretion flow) within each exposure.

\begin{deluxetable*}{lcccccc}
\tablecaption{\suzaku\ Observations \label{suzakuobs}}
\tablewidth{5.3in} \tablehead{ \colhead{}& \colhead{}& \colhead{}&
  \colhead{}& \multicolumn{3}{c}{Clean Exposure (ks)\tnm{b}}
  \\ \colhead{Obs}& \colhead{ObsID} & \colhead{Start Date} &
  \colhead{$\phi_{\rm orb}$\tnm{a}} & \colhead{XIS} & \colhead{PIN} & \colhead{Merge\tnm{c}}}
\startdata 
1 & 702038010 & 2008 Jan 15 08:39:26 & 0.46--0.78 & 21.7 & 22.9 & 20.5 \\ 
2 & 702037010 & 2008 Feb 11 00:42:25 & 0.42--0.80 & 23.4 & 20.0 & 17.9 \\ 
3 & 702036020 & 2008 Apr 5  13:15:26 & 0.11--0.64 & 25.3 & 24.5 & 21.4 \enddata
\tnt{a}{Based on the orbital ephemeris of \citet{levi00}.}
\tnt{b}{Note that because of gaps in the exposures, the clean exposure
  times are roughly half the total duration of each observation as
  indicated by the range of orbital phases.}
\tnt{c}{Total good exposure time used for spectral analysis, obtained by merging
the good time intervals of the XIS, PIN, and PIN background event files.}
\end{deluxetable*}


\subsubsection{XIS}

XIS consists of four detectors, each with $1024\times 1024$ pixel
X-ray-sensitive CCDs at the foci of each of the four X-ray Telescopes
 \citep{naik08}.  However, at the time when the data were taken,
only three of the XIS detectors (XIS0, XIS1, and XIS3) worked
normally.  Of these, XIS0 and XIS3 are front-illuminated (FI) chips,
while XIS1 is a back-illuminated chip.  XIS1 has a significantly
different spectral response from the FI chips, with greater
sensitivity at energies below $\sim$1.5 keV and poorer sensitivity
above $\sim$ 3 keV.  All XIS chips have an energy resolution $\sim130$
eV (FWHM) at 6 keV \citep{mits07}.  The XIS was operated with the
``1/8 Window'' option, which gave a time resolution of 1 s, covering a
field of view of $17\arcmin.8 \times 2\arcmin.2$.

\begin{figure}
\plotone{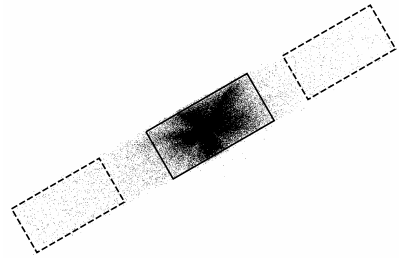}
\caption{Source and background regions for analysis of \suzaku\
  observations of LMC X-4.  This image shows chip XI0 for observation
  1.  To subtract background, we extracted spectra from the source
  region (solid) and background regions (dashed).   \label{reg}}
\end{figure}

We used ds9 V5.6.3, an astronomical imaging and data visualization
application developed by SAO, XSELECT V2.4a, and XSPEC V12.6.0 for
data analysis.  We used XSELECT to
combine XIS exposures taken in $3\times3$ and $5\times5$ event modes to
make single event files.  Next, we defined the regions for the source
and the background for each observation, as shown in Figure~ \ref{reg}.
The source region was a rectangle of size $300''\times140''$, centered
on the source and tilted 30\dgr\ (for XIS0 and XIS3) and 119\dgr\ (for
XIS1).  Background events were extracted from two rectangles of size
$260''\times130''$ on either side of the source and aligned with the
source region.  We note that the XIS background is extremely low, so
that background contributes only $\approx$0.5\% of the total 0.5--10
keV flux from the source.

For spectral analysis, we filtered the XIS events using a good time
interval (GTI) created by merging GTIs for the cleaned XIS and PIN
event files as well as the PIN background event file (described in
Section~\ref{pin}).  The merged clean exposures range from $\approx$18 to
21 ks (Table \ref{suzakuobs}).  Using the filtered XIS event files, we
extracted spectra from the source and background regions and grouped
the spectra so that the total number of events in each energy interval
was roughly equal.  The energy bands below 2.5 keV, 2.5--7 keV, and above
7 keV are grouped by factors of 8, 16, and 32, respectively. Each
(ungrouped) channel has a width of 3.65 eV.

We used the tools {\tt xisrmfgen} and {\tt xissimarfgen} available in
the HEASOFT version 6.9 to generate XIS redistribution matrix files and
ancillary response files.  Following the
recommendations in the \suzaku\ ABC guide\footnotemark, we generated
400,000 simulated photons in creating each ancillary response.  For
each observation, we combined the data from the two FI chips (XIS0 and XIS3)
into a single spectrum.

\footnotetext{\suzaku\ ABC guide, v. 2, Section~6.6.4;
JX-ISAS-SUZAKU-MEMO-2007-04.}

We note that ``wobble'' of the {\em Suzaku} spacecraft, owing to
time-dependent thermal distortions of its chassis, can cause the
pointing position to vary by as much as $\sim$40\arcsec.  As this is a
significant fraction of the size of the 1/8 Window (our extraction
regions are 140\arcsec\ wide), wobble may cause source flux to be lost
outside the window during the XIS exposures.  We examine the thermal
wobble by deriving the centroid position of the source in detector
coordinates ({\tt DETX} and {\tt DETY} in the event files) in bins of
size 50 s.  For $>97\%$ of the total exposure time the thermal wobble
is $<$20\arcsec, for which there is negligible flux loss.  Observation
3 has a brief period ($<$2 ks) of wobble $>$30\arcsec.  We derive a
rough estimate of the flux lost outside the window during these
periods, by calculating the observed source counts after shifting the
extraction region by an amount and direction corresponding to the
observed thermal wobble.  At most only 15\% of the flux is lost
outside the window at any given time, and the total flux lost over the
full observation is $<$1.5\%.  We thus verify that losses due to
thermal wobble are minimal, and for simplicity we do not exclude the
brief intervals of large wobble from the analysis.

\footnotetext{JX-ISAS-SUZAKU-MEMO-2007-04}

\subsubsection{PIN}

\label{pin}

We include PIN data in the spectral and timing analysis.  As the
HXD is a non-imaging instrument, we cannot estimate background
directly from the observations.  The PIN background consists of an
instrumental component as well as the cosmic X-ray background (CXB).
The instrumental background is estimated using the version 2 ``tuned''
background event files\footnotemark.  Spectra were extracted for both
the observations and the background file and corrected for dead time.
As with the XIS, spectra were extracted using the merged GTI described
above.
\footnotetext{ftp://legacy.gsfc.nasa.gov/suzaku/data/background/pinnxb\_ver2.0/}
The CXB component is modeled during spectral analysis, as a power law
with a high-energy exponential cutoff\footnotemark; this component is
only $\sim$2\% of the observed flux.

\footnotetext{\suzaku\ ABC guide, v. 2, Section~7.3.3.1}


\subsection{{\it RXTE}}

\rxte\ was launched on 1995 December 30 \citep{swan06}.  The
\rxte\ ASM observes bright X-ray sources in the range of 1.5 -- 12
keV, to explore the variability of sources including black holes,
neutron stars, X-ray pulsars, and bursts of X-rays.  ASM scans about
80\% of the sky every orbit, allowing monitoring on timescales of 90
minutes or longer.  Details about the \rxte\ ASM can be found in
\citet{levi96}.  In this paper, we use 13 years of the \rxte\ ASM data
on LMC X-4 from 1996 to 2009.  We retrieve the one-day average
light curve from the public MIT
Web site\footnotemark\footnotetext{http://xte.mit.edu/ASM\_lc.html}.


\section{Data Analysis and Results}
\label{analysis}

\subsection{Superorbital Period Determination}

\begin{figure*}
\plotone{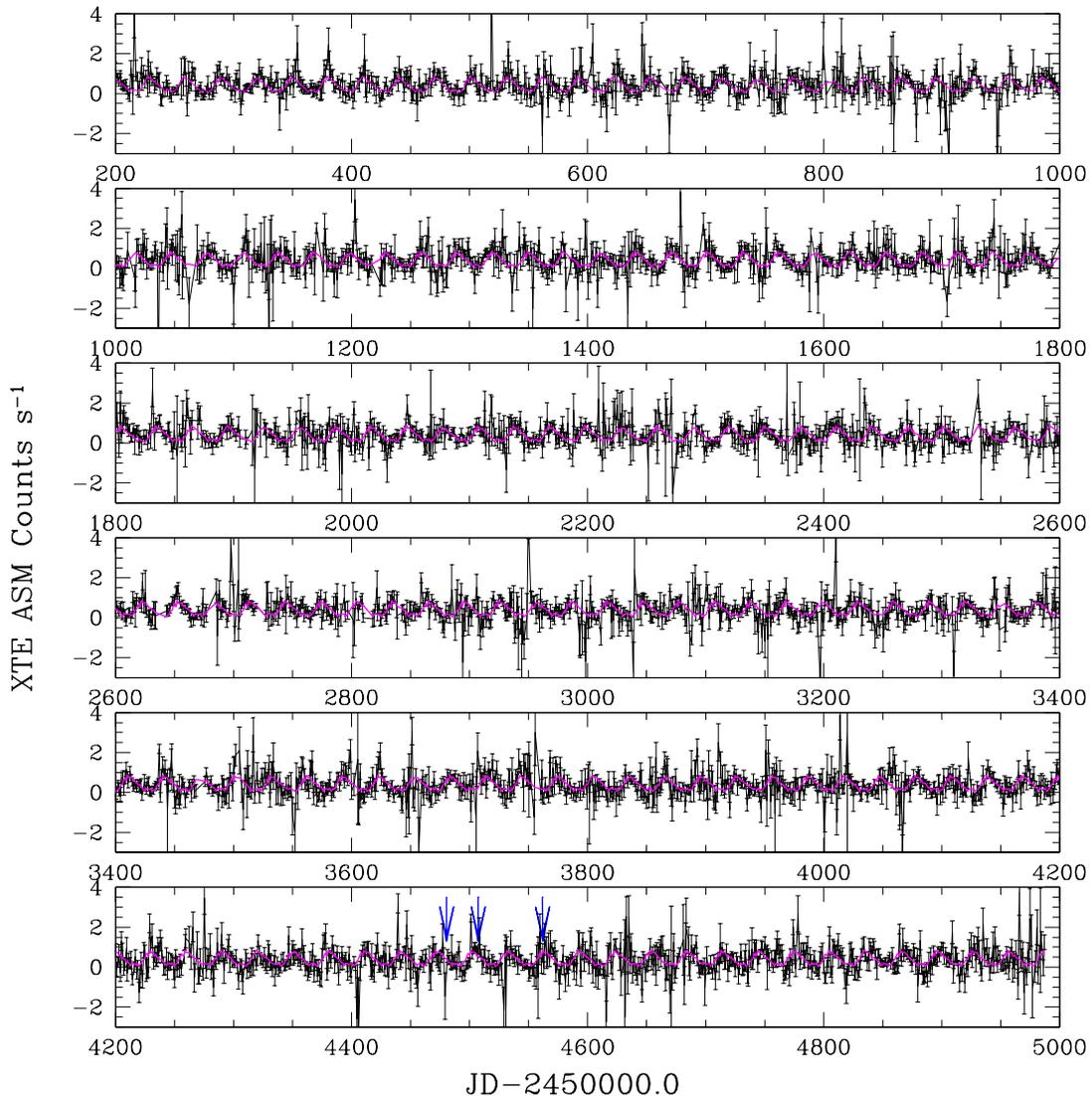}
\caption{One-day average light curve of LMC X-4 covering 13 years of
  observations (1996 -- 2009) by the \rxte\ ASM.  The magenta line is
  our best fit to the $30.32\pm0.04$ days superorbital period of LMC
  X-4. The hats on the error bars denote the widths of the time over
  which the data were averaged (1 day for the ASM
  data). \vskip1ex \label{1day}}
\end{figure*}

In order to determine the superorbital period during the time of our
Suzaku observations, we first apply an epoch folding method to the 13
years of \rxte\ data to obtain an updated superorbital period.  By
comparing the variance of the folded light curve
with the variance within each bin \citep[the ANOVA
method;][]{davi90anova,davi91anova}, we calculate the Davies L
statistic for different superorbital periods.  Gaussian fits to the
period versus L statistic, for different numbers of phase bins, all give
about $30.316$ days at high significance.  A power spectrum using the
Lomb Normalized Periodogram \citep{scar82} is also consistent with
30.32 days.  This result is consistent with the superorbital period
determined using five years of \rxte\ data by \citet{paul02a}.  This
suggests no significant change in the long-term period over the 13
years of \rxte\ observations, in contrast to the results of
\citet{paul02a} who had found a long-term decrease in the period over
the preceding ~20 years.  We estimate the error by re-folding the
light curve into
various numbers of bins and examining the resulting best fit.
Both the ANOVA and Lomb Normalized Periodogram methods determine a
long-term period of $30.32\pm0.04$ days.  Figure~\ref{1day} shows a plot
of our best-fit period superposed on the 13 year \rxte\ ASM light curve.
A study was made of possible variability in the long-term period over
the \rxte\ ASM lifetime and the folded light curve gave similar
results near the times of the \suzaku\ observations as the steady
period model.  

We note that the L statistic does not make use of uncertainties in the
\rxte\ lightcurve.  As a check, we performed an $e$-folding analysis
using the FTOOL {\tt efsearch} to derive the period that maximizes
$\chi^2$ relative to no variation.  We used 8, 16, or 32 phase bins
with time resolutions of 0.09, 0.05, or 0.02 days, respectively.  We
obtained long-term periods of $30.29\pm0.09$, $30.29\pm0.05$,
and $30.32\pm0.02$ (where the errors are the period resolutions of the
search), corresponding to maximum $\chi^2$ of 890, 959, and 1070,
respectively.  These estimates are consistent with our result derived
from the ANOVA and Lomb Normalized Periodogram methods.

\begin{figure*}
\plotone{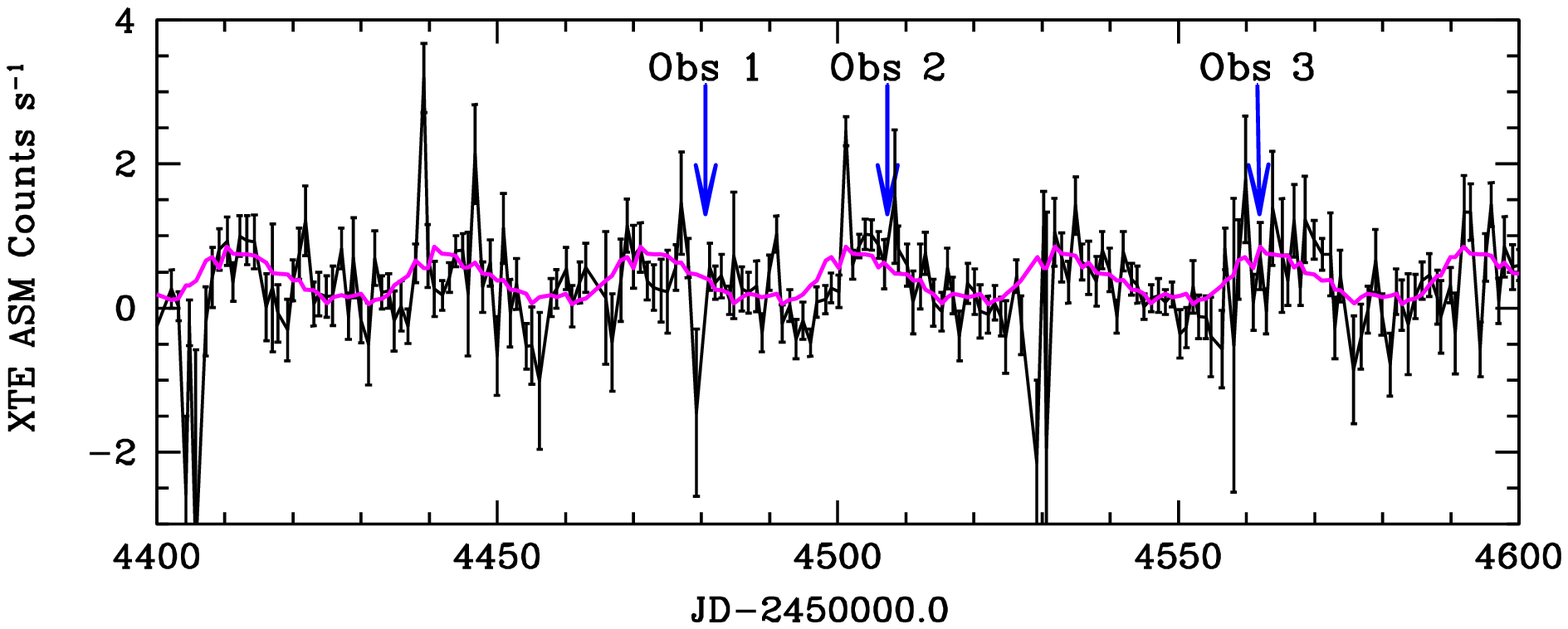}
\caption{Close up of ASM count and the region of the
  \suzaku\ observations.  Blue arrows indicate the times of our
  \suzaku\ observations.  The superorbital phases for the three
  observations (left to right) are $\phi_{30} =0.39$, $\phi_{30} =
  0.27$, and $\phi_{30} = 0.07$.  The zero superorbital phase is defined
  as the beginning of a high state (JD 2454560.0).  The magenta line
  is our best fit of the superorbital period of LMC X-4. The hats on
  the error bars denote the widths of the time over which the data
  were averaged (1 day for the ASM data). The widths of the arrows
  denote the length of our observations (about 20ks each).
\label{lc}}
\end{figure*}

Our improved fit to the superorbital period allows us
to accurately determine the phases of our \suzaku\ observations as
shown in Figure~\ref{lc}.  We define the start of the high state of the
superorbital cycle ($\phi_{30} = 0$) to be at JD 2454560.0.

\subsection{Phase-averaged Spectroscopy}
\label{subsec:spectroscopy}

For each observation, we perform spectroscopic fits to the XIS FI, XIS
1, and PIN spectra simultaneously, for a combined spectral range of
0.6--50 keV.  We fit all three spectra with the same spectral model,
but allow the overall normalizations to vary in order to allow for
uncertainties or systematic offsets in the flux
calibration.  Fits to the power-law spectrum of the Crab
nebula indicate that the best-fit normalizations for the three XIS
chips are consistent to within a few percent, while the best-fit
normalization for the PIN relative to XIS is 1.16\footnotemark.

\footnotetext{
  JX-ISAS-SUZAKU-MEMO-2008-06}

We first fit the phase-averaged spectra with a simple continuum model
of power-law plus blackbody that has previously been successfully
employed for LMC X-4 \citep{paul02}, as well as a number of other
X-ray pulsars \citep{hick04}.  Previous studies have found evidence
for a high-energy cutoff in the power-law component; \citet{laba01}
use a cutoff of the form
\begin{equation}
e^{(E-E_{\rm cut})/E_{\rm fold}}; E\ge E_{\rm cut}
\end{equation}
with 
 $E_{\rm
  cut}\approx E_{\rm fold} \approx 18$ keV.  We find that for the {\em Suzaku} data the spectra are better described by a Fermi-Dirac cutoff \citep[e.g.,][]{cobu02xbp,rive10xbp} of the form:

\begin{equation}
1/(1+e^{(E-E_{\rm cut})/E_{\rm fold}}),
\end{equation}
with $E_{\rm cut}\approx 27$ keV and $E_{\rm fold}\approx 8$ keV.

\begin{figure}
\plotone{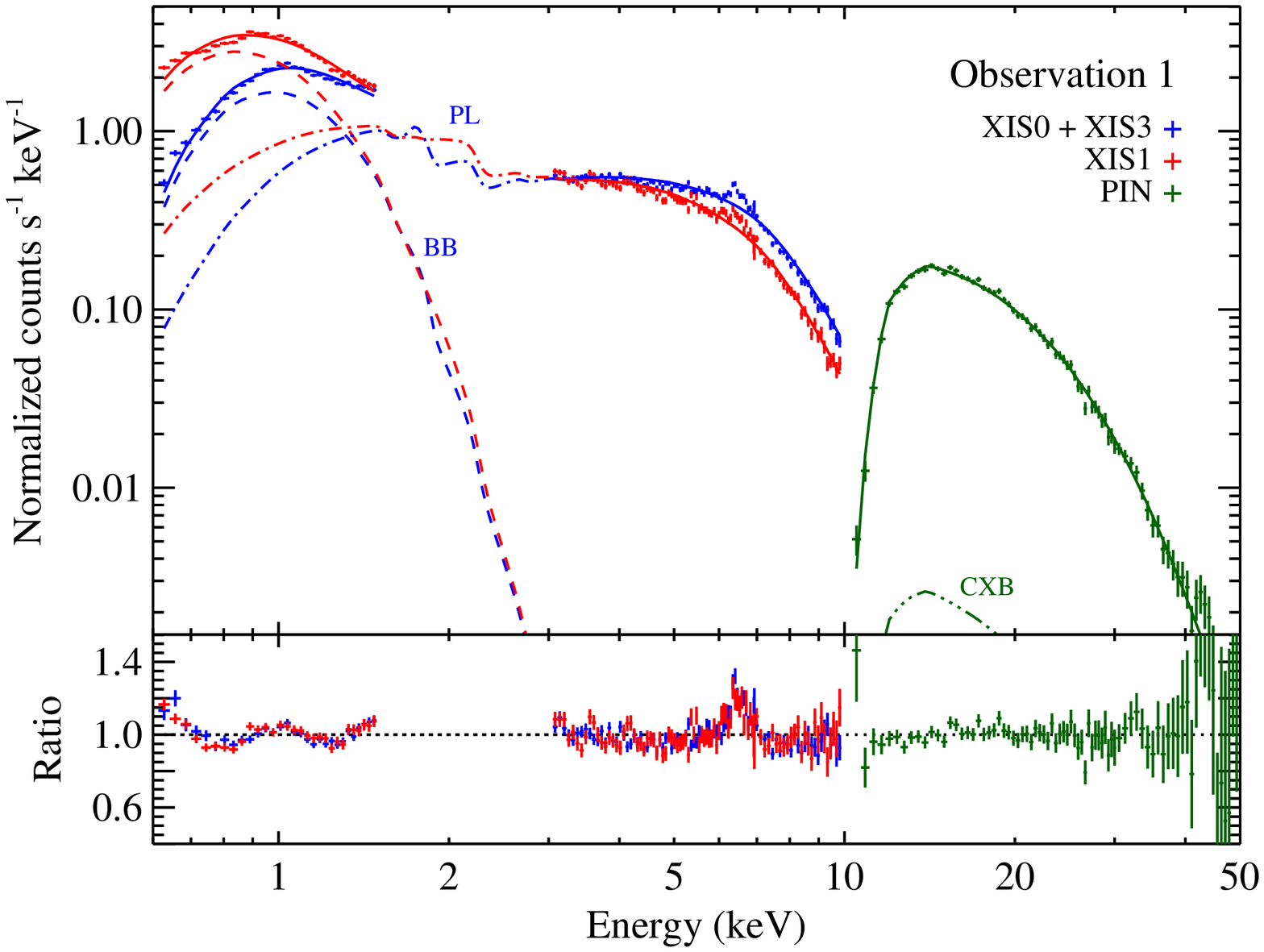}\\
\plotone{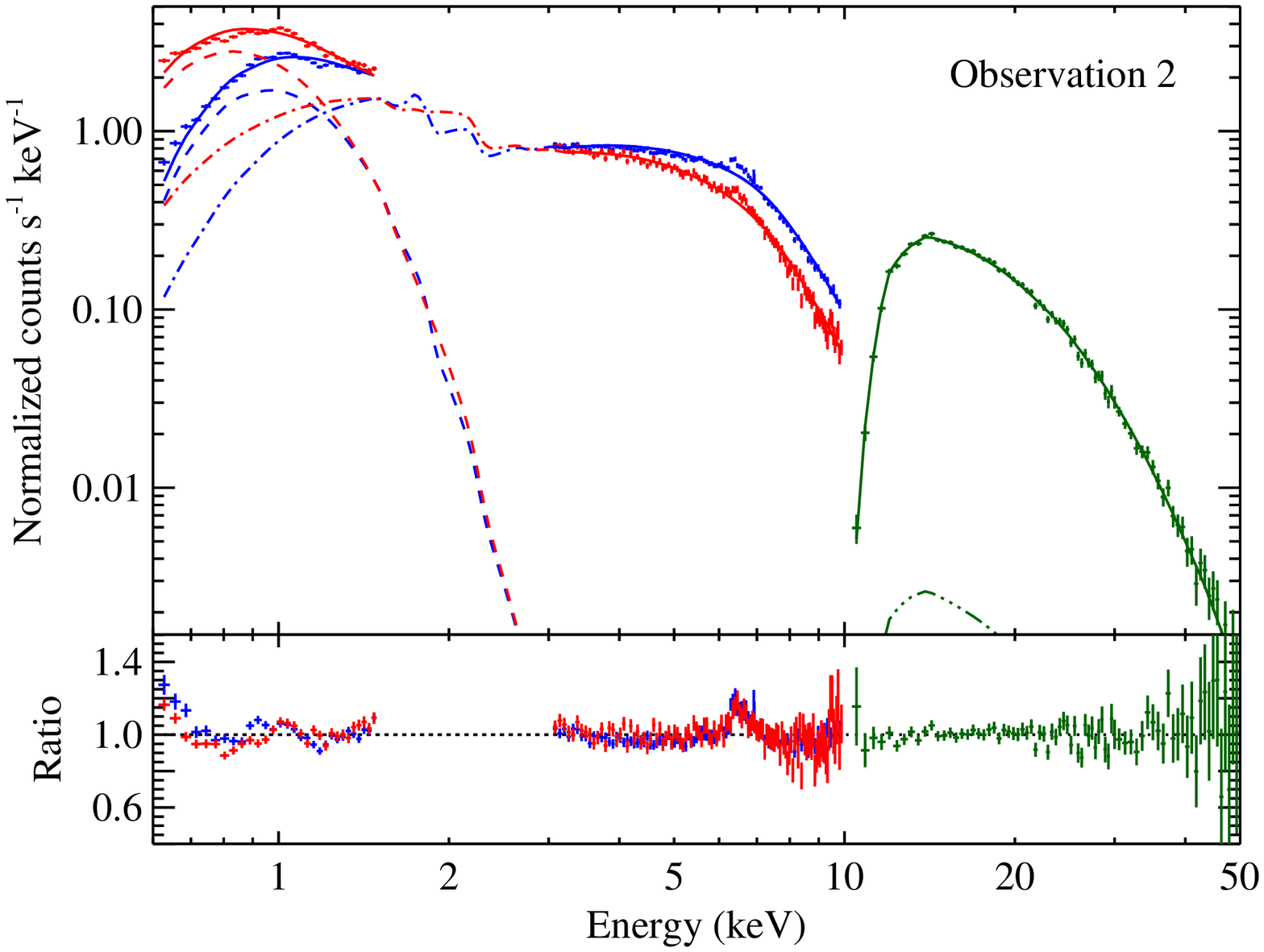}\\
\plotone{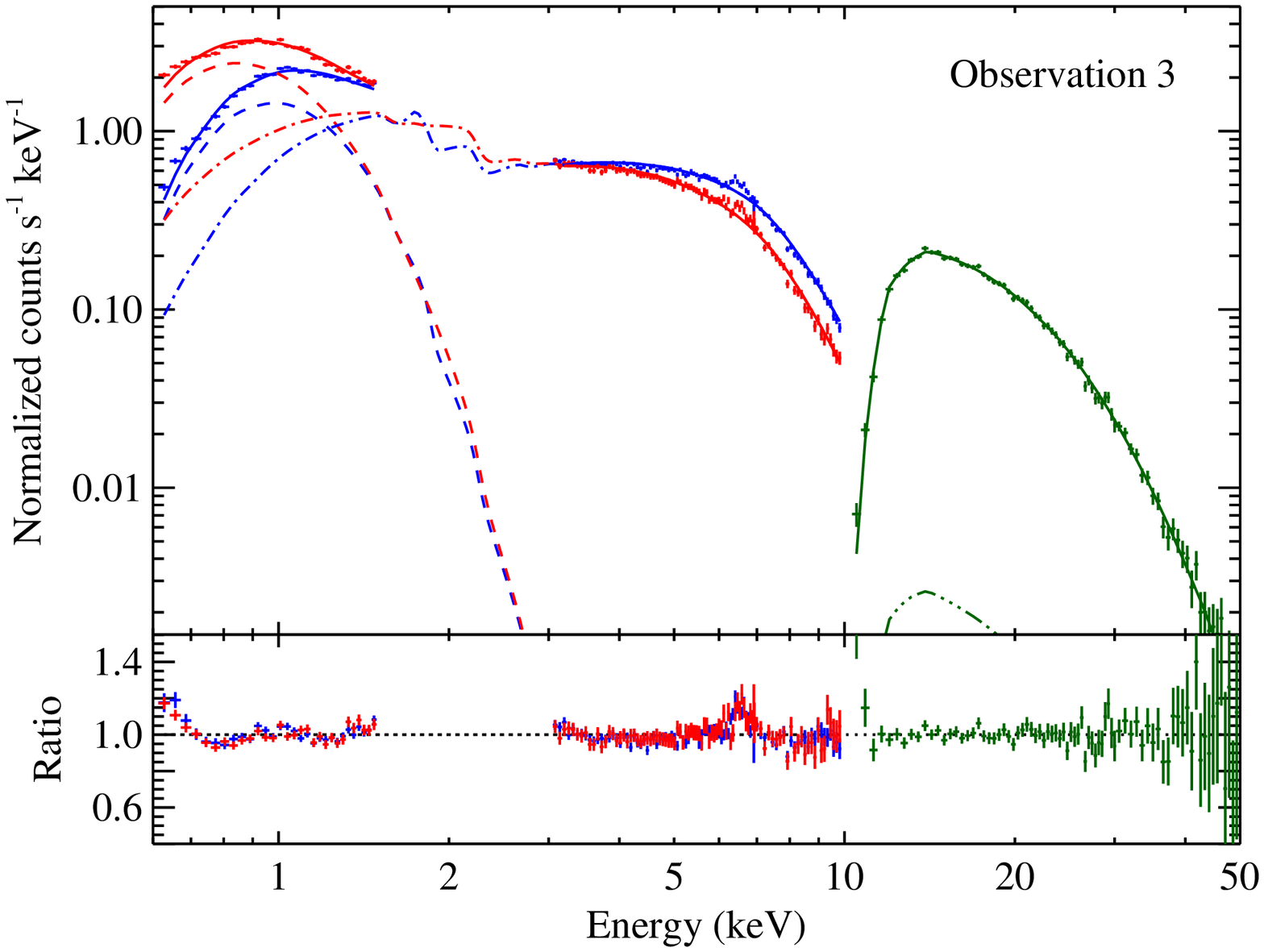}\\
\caption{Phase-averaged \suzaku\ spectra for the three observations,
  fitted with a continuum consisting of a blackbody (dashed) plus
  power law with high-energy cutoff (dash dot).  The FI, XIS1, and PIN
  spectra are shown in blue, red, and green, respectively.  The
  individual spectral components are shown, including the CXB model
  for the PIN spectrum.  Note the significant line-like
  residuals. \label{fitbbpo}}
\end{figure}

In the XIS data, we exclude
the energy range 1.5--3 keV to avoid instrumental effects near the
Au and Si edges \citep{mill08,saez09}. We include neutral hydrogen
absorption fixed to the Galactic value of $N_{\rm H}=0.057 \times
10^{22}\ \rm cm^{-2}$ \citep{paul02}.  Our best-fit parameters for
this simple model give a blackbody with $kT_{\rm BB} \sim 0.18$ keV
and a power law with photon index $\Gamma \sim 0.6$, similar to
results from previous studies \citep{paul02,laba01}.  Spectral fits
are shown in Figure\ \ref{fitbbpo}, and fit parameters are given in
Table\ \ref{finalfitbbpo}.

A single cutoff power law provides a reasonable fit to the high-energy
component in the full range from 1 to 50 keV; the best-fit photon index
for the joint XIS and PIN analysis is consistent (within 2\%) from
that derived from fitting the XIS data alone.  We note that
Figure~\ref{fitbbpo} shows a departure from the cutoff power law at
$\sim$40 keV for observation 1.  This feature does not appear in
observations 2 and 3, and so for simplicity we model all three
observations with the simple cutoff power law.

  \epsscale{1.1}

\begin{deluxetable*}{lcccc}
\tablecaption{Fits Using a Power Law and a Blackbody to \suzaku\ Observations \label{finalfitbbpo}}
\tablewidth{6.3in}
\tablehead{
\colhead{Parameter}&
\colhead{Observation 1} &
\colhead{Observation 2} &
\colhead{Observation 3}}
\startdata
$\chi^2_{\nu}$ (317 dof) & 2.90 & 2.53 & 2.18  \\
$N_{\rm H}$ ($\times 10^{22}\ \rm cm^{-2}$) (fixed) & $5.7 \times10^{-2}$ & $5.7 \times10^{-2}$ & $5.7 \times10^{-2}$ \\
\cutinhead{Continuum}
$kT_{\rm BB}$ (keV) & $0.181\pm0.002$ & $0.174\pm0.002$ & $0.181\pm0.002$ \\
$N_{\rm BB}$\tnm{a} & $(6.32^{+0.09}_{-0.08}) \times10^{-4}$ & $(6.99\pm0.11) \times10^{-4}$ & $(5.57^{+0.09}_{-0.11}) \times10^{-4}$ \\
Power-law $\rm \Gamma$ & $0.56\pm0.02$ & $0.56\pm0.02$ & $0.56\pm0.02$ \\
$N_{\rm \Gamma}$\tnm{b} & $(0.67\pm0.02) \times10^{-2}$ & $(1.05\pm0.03) \times10^{-2}$ & $(0.83\pm0.02) \times10^{-2}$ \\
$E_{\rm cut}$ (keV) & $25.3^{+ 1.0}_{- 1.2}$ & $24.8^{+ 1.3}_{- 1.6}$ & $24.7^{+ 1.2}_{- 1.5}$ \\
$E_{\rm fold}$ (keV) & $ 7.1^{+ 0.6}_{- 0.5}$ & $ 8.6\pm 0.6$ & $ 8.1\pm 0.6$ \\
\cutinhead{Relative Normalizations}
$A_{\rm XIS1}$ & $1.01\pm0.01$ & $0.95\pm0.01$ & $1.02\pm0.01$ \\
$A_{\rm PIN}$ & $1.07\pm0.03$ & $1.05\pm0.03$ & $1.09\pm0.03$
\enddata
\tnt{a}{XSPEC normalization, units of $8.36\times10^{-8}$ \flux\ (bolometric).}
\tnt{b}{XSPEC normalization, units of photons cm$^{-2}$ s$^{-1}$ at 1 keV.}

\end{deluxetable*}

\begin{figure}
\plotone{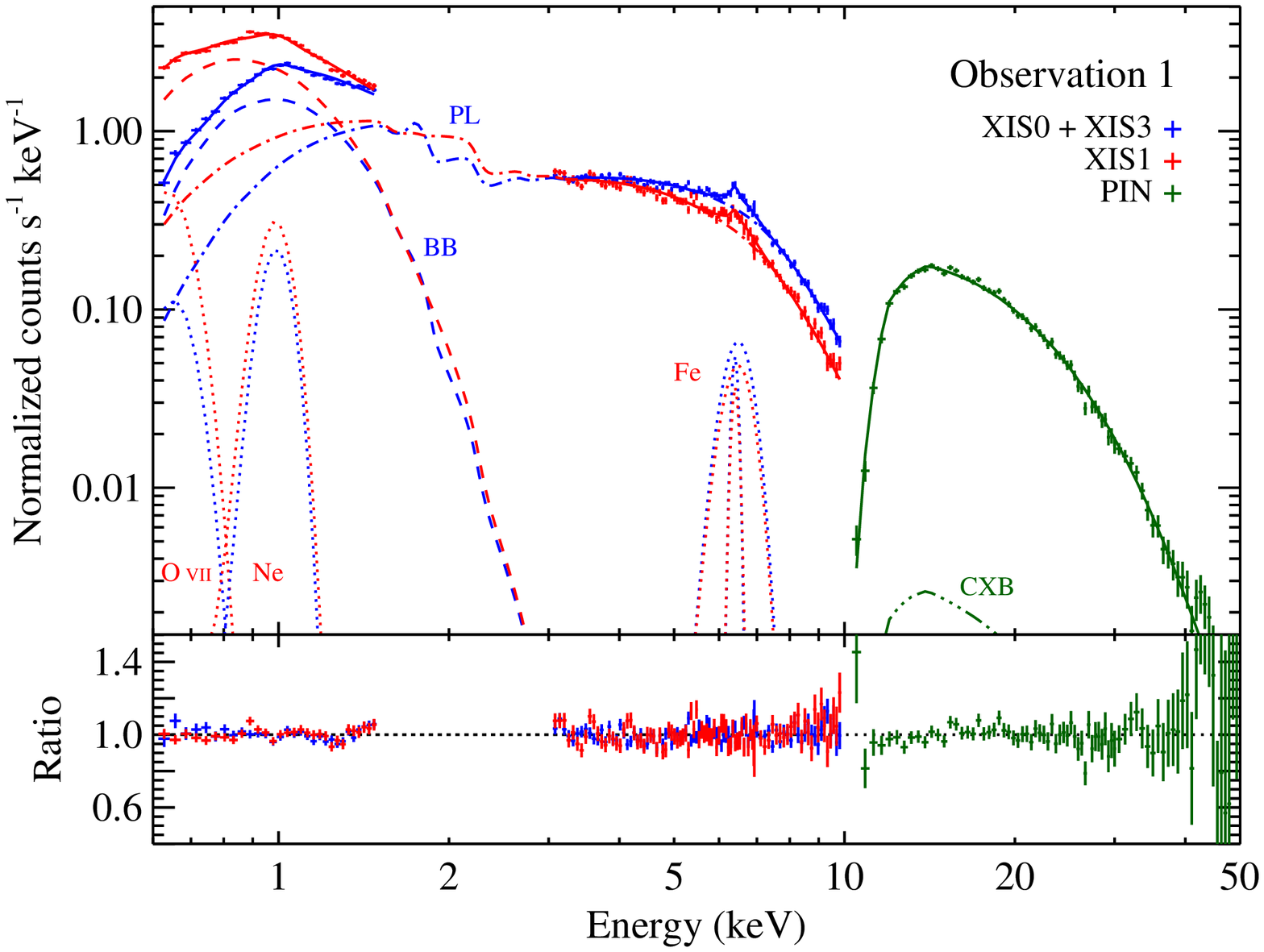}\\
\plotone{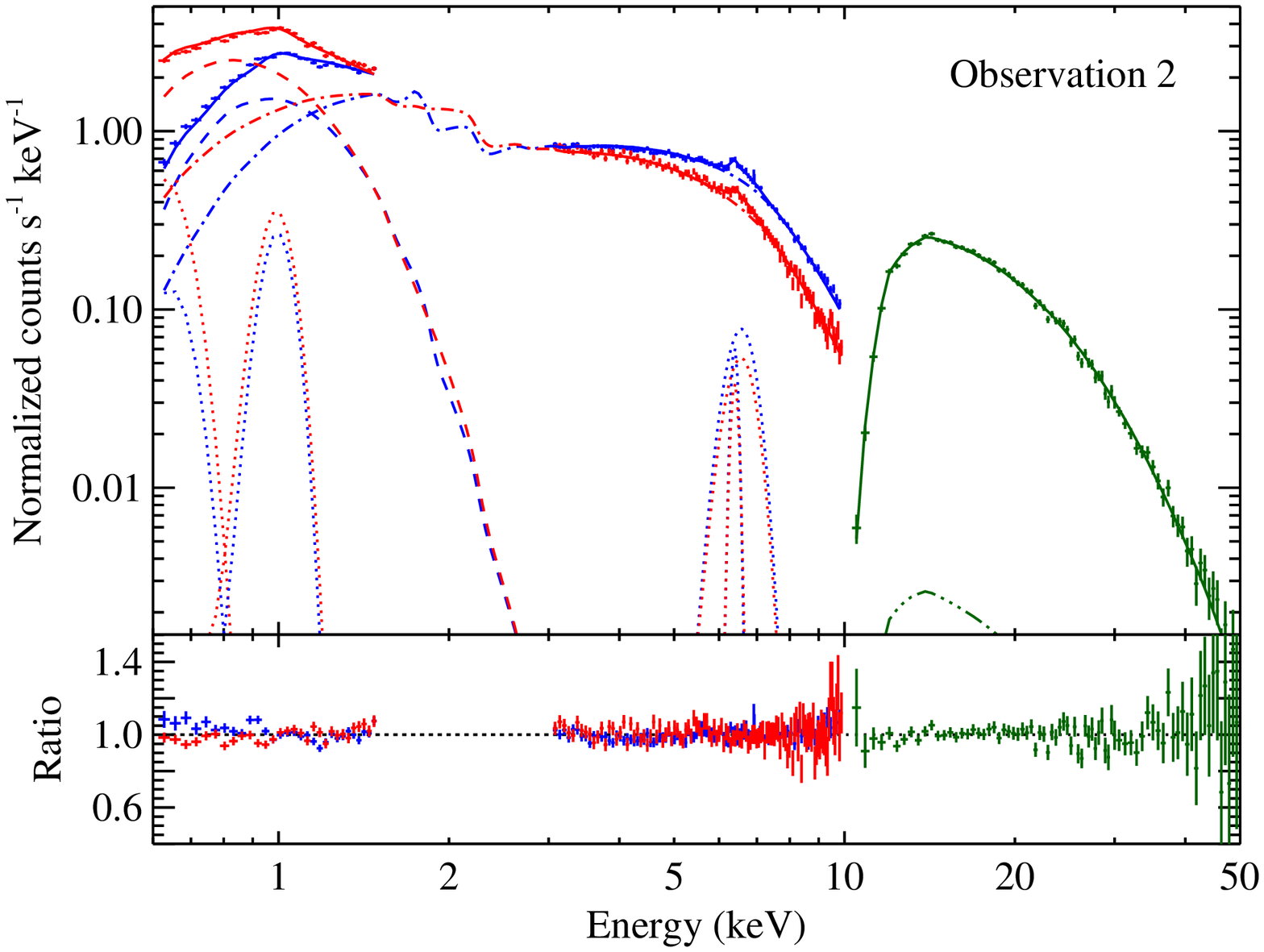}\\
\plotone{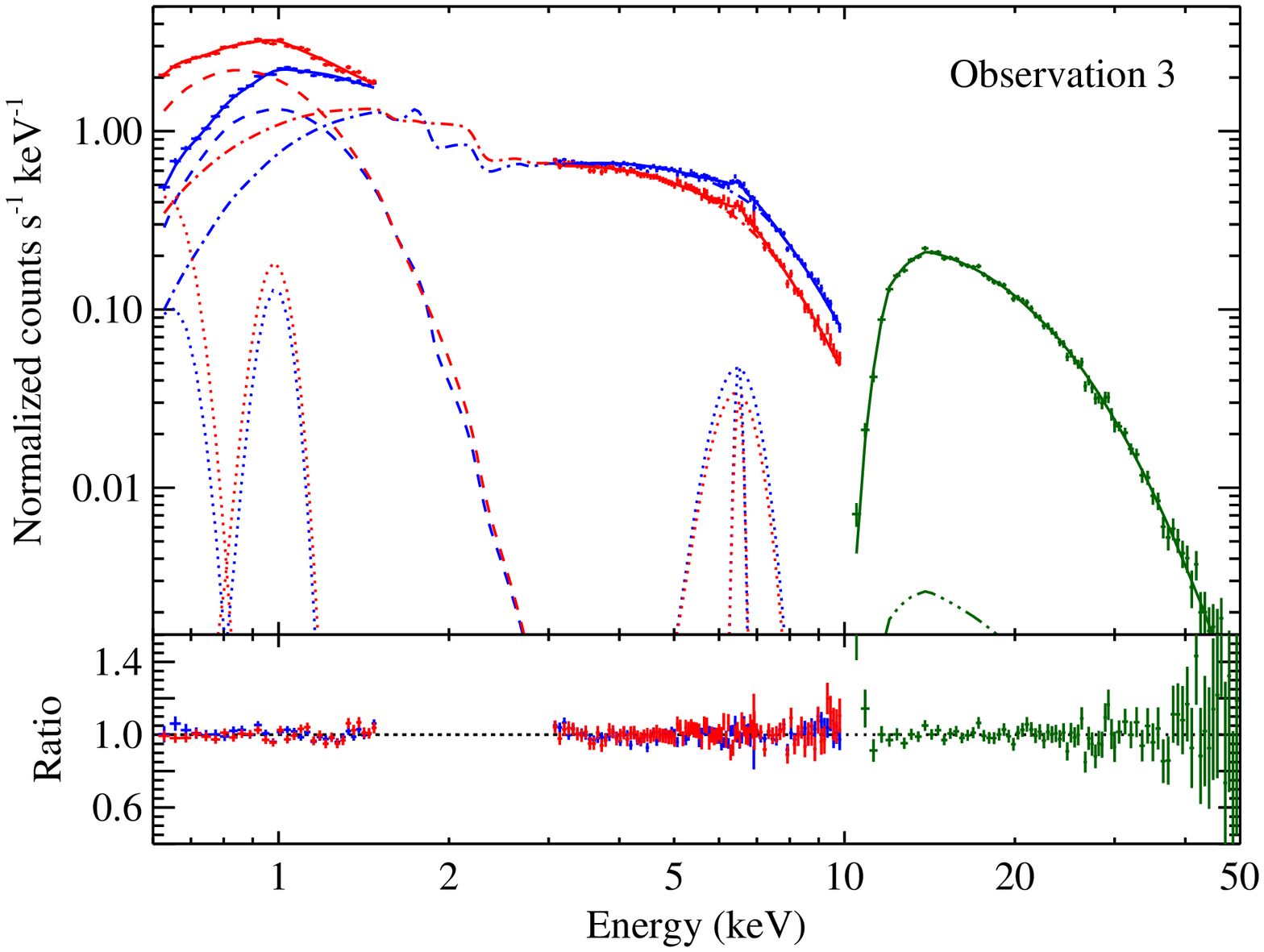} \\
\caption{Phase-averaged spectra from \suzaku\ observations with
  blackbody (dashed), power law with high-energy cutoff (dash dot), and emission line (dotted)
  components shown.  The FI, XIS1, and PIN
  spectra are shown in blue, red, and green, respectively.
The emission lines are broad Fe, Fe K$_\alpha$, O\,{\sc viii}, and Ne\,{\sc ix} ({\sc x} Ly$\alpha$).  The inclusion of the lines significantly
  improves the residuals compared to the continuum fits shown in Figure\
  \ref{fitbbpo}. \vskip1ex \label{fit}}
\end{figure}

In the XIS spectrum at $E<10$ keV, it is clear from the residuals in
Figure~\ref{fitbbpo} that other, line-like spectral features are present
in addition to the simple continuum model.  Emission line features are
often observed in X-ray binaries; for example, in SMC X-1, several
emission lines from O, Ne, Mg, Si, S, Ar, and Fe were needed to fit
the spectrum \citep{vrti05}.  In addition, using the
\chandra\ High-Energy Transmission Grating Spectrometer and the
\xmmnewton\ Reflection Grating Spectrometer, \citet{neil09} performed
X-ray spectroscopy of LMC X-4 and detected several X-ray emission
lines.  In the \suzaku\ observations, we find at least four emission
lines including O\,{\sc viii}, Ne\,{\sc ix} ({\sc x} Ly$_\alpha$), Fe K$_\alpha$, and
broad Fe (line energies and equivalent widths (EWs) are given in
Table\ \ref{lines}).  The total $\chi ^{2}$ values for every
observation are lower by at least 30 when adding each individual
emission line.  Following \citet{avni76} and \citet{yaqo98}, we are
$>$90\% confident that the line features are detected.  The average
$\chi^2$ per degree of freedom for the three observations is reduced
from 2.5 to 1.2. The spectral fits are shown in Figure~\ref{fit}, and
the results are listed in Table\ \ref{finalfit}.

We note that for observations 1 and 3, with the inclusion of emission
lines, the best-fit relative normalizations between the PIN and the XIS
FI spectra are broadly consistent with the value of 1.16 derived from
calibration studies and higher than the value of $\approx$1.07 for
the pure continuum model.

\begin{deluxetable*}{lccccc}
\tablecaption{X-ray Emission Lines Used in Fitting Spectra of LMC X-4 \label{lines}}
\tablewidth{6.3in}
\tablehead{
\multicolumn{1}{c}{} &
\colhead{}&
\colhead{Line}&
\colhead{Observed}&
\colhead{Equivalent} \\
\colhead{Line}&
\colhead{Wavelength (\AA)} &
\colhead{Energy (keV)} &
\colhead{Energy (keV)\tnm{a}} &
\colhead{Width (eV)}\tnm{b}}

\startdata
O\,{\sc viii} Ly$_\alpha$ & 18.97 & 0.65 &0.60--0.62                         & $ 41\pm 13$ \\
Ne\,{\sc ix} (Ne\,{\sc x} Ly$_\alpha$) & 13.45 (12.13)  & 0.91 (1.02) &0.98--0.99   & $ 13\pm  5$ \\
Fe K$_\alpha$ & 1.94 & 6.38 & 6.40--6.51                              & $ 25\pm 19$ \\
Broad Fe & 1.89&\nodata & 6.52--6.61                                  & $150\pm 50$ 
\enddata

\tnt{a}{Range of best-fit line energies for the three observations.}
\tnt{b}{Average EWs and uncertainties between the three observations.}

\end{deluxetable*}

\begin{deluxetable*}{lcccc}
\tabletypesize{\footnotesize}
\tablecaption{Fits Including Emission Lines to \suzaku\ Observations\label{finalfit}}
\tablewidth{6.6in} 
\tablehead{
\colhead{Parameter}&
\colhead{Observation 1} &
\colhead{Observation 2} &
\colhead{Observation 3}}
\startdata
$\chi^2_{\nu}$ (308 dof) & 1.27 & 1.20 & 1.08  \\
$N_{\rm H}$ ($\times 10^{22}\ \rm cm^{-2}$) (fixed) & $5.7 \times10^{-2}$ & $5.7 \times10^{-2}$ & $5.7 \times10^{-2}$ \\
\cutinhead{Continuum}
$kT_{\rm BB}$ (keV) & $0.183^{+0.003}_{-0.002}$ & $0.174\pm0.003$ & $0.183\pm0.003$ \\
$N_{\rm BB}$\tnm{a} & $(5.69^{+0.12}_{-0.13}) \times10^{-4}$ & $(6.24^{+0.16}_{-0.13}) \times10^{-4}$ & $(5.05\pm0.12) \times10^{-4}$ \\
Power-law $\rm \Gamma$ & $0.64\pm0.02$ & $0.63\pm0.02$ & $0.62\pm0.02$ \\
$N_{\rm \Gamma}$\tnm{b} & $(0.73\pm0.02) \times10^{-2}$ & $(1.12\pm0.03) \times10^{-2}$ & $(0.88^{+0.03}_{-0.02}) \times10^{-2}$ \\
$E_{\rm cut}$ (keV) & $26.7^{+ 0.9}_{- 1.0}$ & $26.5^{+ 1.2}_{- 1.3}$ & $26.1^{+ 1.1}_{- 1.3}$ \\
$E_{\rm fold}$ (keV) & $ 6.8^{+ 0.6}_{- 0.5}$ & $ 8.3\pm 0.6$ & $ 7.9\pm 0.6$ \\
\cutinhead{Emission Lines}
$E_{\rm O}$ (keV) & $0.62\pm0.02$ & $0.60\pm0.02$ & $0.61\pm0.02$ \\
$\sigma_{\rm O}$ (keV) & $5.0 \times10^{-2}$ & $5.0 \times10^{-2}$ & $5.0 \times10^{-2}$ \\
$I_{\rm O}$ (photons cm$^{-2} \rm\ s^{-1}$) & $(2.5^{+0.7}_{-0.5}) \times10^{-3}$ & $(3.4^{+1.6}_{-0.8}) \times10^{-3}$ & $(2.5^{+0.7}_{-0.6}) \times10^{-3}$ \\
$E_{\rm Ne}$ (keV) & $0.98\pm0.02$ & $0.99^{+0.02}_{-0.01}$ & $0.98\pm0.03$ \\
$\sigma_{\rm Ne}$ (keV) & $5.0 \times10^{-2}$ & $5.0 \times10^{-2}$ & $5.0 \times10^{-2}$ \\
$I_{\rm Ne}$ (photons cm$^{-2} \rm\ s^{-1}$) & $(3.8^{+0.8}_{-0.7}) \times10^{-4}$ & $(4.7^{+0.8}_{-0.9}) \times10^{-4}$ & $(2.4\pm0.7) \times10^{-4}$ \\
$E_{\rm Fe}$ (keV) first Gaussian & $6.41\pm0.03$ & $6.40\pm0.04$ & $6.51^{+0.06}_{-0.05}$ \\
$\sigma_{\rm Fe}$ (keV) & $5.0 \times10^{-2}$ & $5.0 \times10^{-2}$ & $5.0 \times10^{-2}$ \\
$I_{\rm Fe}$ (photons cm$^{-2} \rm\ s^{-1}$) & $(7.0^{+2.6}_{-2.7}) \times10^{-5}$ & $(7.9^{+3.0}_{-3.2}) \times10^{-5}$ & $(6.5^{+2.3}_{-2.4}) \times10^{-5}$ \\
$E_{\rm Fe}$ (keV) second Gaussian & $6.53^{+0.06}_{-0.05}$ & $6.61^{+0.07}_{-0.06}$ & $6.52\pm0.09$ \\
$\sigma_{\rm Fe}$ (keV) & $0.37^{+0.08}_{-0.07}$ & $0.37^{+0.09}_{-0.07}$ & $0.53^{+0.15}_{-0.14}$ \\
$I_{\rm Fe}$ (photons cm$^{-2} \rm\ s^{-1}$) & $(3.7^{+0.6}_{-0.5}) \times10^{-4}$ & $(4.5^{+0.9}_{-0.8}) \times10^{-4}$ & $(3.7^{+0.5}_{-0.8}) \times10^{-4}$ \\
\cutinhead{Relative Normalizations}
$A_{\rm XIS1}$ & $1.01\pm0.01$ & $0.95\pm0.01$ & $1.01\pm0.01$ \\
$A_{\rm PIN}$ & $1.15^{+0.04}_{-0.03}$ & $1.12\pm0.03$ & $1.15\pm0.03$

\enddata

\tnt{a}{XSPEC normalization, units of $8.36\times10^{-8}$ \flux\ (bolometric).}
\tnt{b}{XSPEC normalization, units of photons cm$^{-2}$ s$^{-1}$ at 1 keV.}

\end{deluxetable*}

As an alternative to blackbody model for the soft component, we also
consider models for optically thin thermal emission, which have
previously been used to describe the soft emission from LMC X-4
\citep{paul02,naik04}.  We consider both a thermal bremsstrahlung
model and an optically thin (MEKAL) model that includes contributions
from metal lines \citep[e.g.,][]{habe05rxj}.  We perform fits using a
model identical to that described above including emission lines, but
replacing the blackbody component with the thermal bremsstrahlung and
MEKAL models.  We find that these can represent the soft component
equally well to the blackbody model (although the MEKAL model requires
very low metallicity $<2$\% of solar).  Therefore, we cannot rule out these
models for the soft excess on spectral grounds.  However, we
can rule them out on physical grounds, following the argument of
\citep{hick04}. It is possible that the diffuse cloud of gas around
the neutron star \citep{boro01} would reprocess the hard X-ray and
produce optically thin emission that might be described by
optically thin emission models.  However, in order to produce high
$L_{\rm soft}$ as observed in LMC X-4, the system must have the
emission region $R_{\rm cloud} > 10^{12}\ \rm cm$, which is larger
than the size of the emission region $R_{\rm cloud} \lesssim
10^{11}\ \rm cm$ as inferred from the decrease in soft flux during the
eclipse.  Further, such a large region cannot produce soft pulsations
as observed.  Therefore, although the optically thin models provide an
adequate description of the data, we prefer the blackbody model for
our physical interpretation of the soft component.

For all the spectral fits shown in Figs.~\ref{fitbbpo} and \ref{fit},
the blackbody component dominates the energy range below 1 keV, while
power-law emission dominates above 2 keV.  We propose that the hard
(power-law) X-rays from the pulsar are reprocessed by the inner
accretion disk, where the soft pulses are emitted.  The power-law flux
is $(8$--$10) \times10^{-9}$ \flux, assuming the absolute flux
calibration of the XIS FI chips.
Taking the distance to LMC to be 50 kpc, we obtain $L_{X}\sim 3 \times
10^{38}$ \ergs\ from 2 to 50 keV.  The bolometric luminosity of the
blackbody component (taken directly from the XSPEC normalization) is
$\sim1.5\times10^{37}$ \ergs.

Using the observed temperatures and luminosities, we can derive the
size of the blackbody-emitting region, assuming that the emission is
from a spherical region that partially covers the central source.
Note that this differs from another commonly used estimate for the
blackbody radius, which assumes a circular emission geometry.  We
prefer the spherical partial covering geometry, because for this source the
blackbody emission is likely due to the reprocessing of the hard
emission from the neutron star \citep{hick04}.  For a partial covering spherical shell and an incident hard X-ray
luminosity $L_X$, the radius of the shell (as shown in Figure~9 of
\citealt{hick04}) is
\beq 
R_{\rm BB}^2 =
{L_{X}\over 4\pi \sigma T_{\rm BB}^4}.
\eeq 
Using $L_{X}\sim 3\times 10^{38}$ \ergs\ and $kT_{\rm BB} \sim
0.18\ \rm keV$, we have $R_{\rm BB}\sim 1.5\times 10^{8}\ \rm cm$.  In
this picture, the covering fraction $\Omega$ of the spherical shell is given by
simply $\Omega = L_{\rm BB}/L_X$ and so is relatively small ($\Omega \sim 0.05$).

 In LMC X-4, an estimate for the magnetic field comes from the
 observed cyclotron feature at $\sim$ 100 keV \citep{laba01}.  We note
 that the existence of this feature has been a matter of debate
 \citep[e.g.,][]{tsyg05lmcx4}, but since it provides the only existing
 estimate for the magnetic field in this source, we adopt it in the
 subsequent analysis.  A cyclotron energy of 100 keV would indicate $B
 \sim 10^{13}$ G.  An estimate of the magnetospheric radius \citep[e.g.,][]{ghos79torque, fran02}
is 
\begin{equation}
R_{\rm m}\sim0.5
R_{\rm A}\sim1.5\times10^{8}m_1^{1/7}R_6^{10/7}L_{37}^{-2/7}B_{12}^{4/7} {\rm
cm,}
\end{equation}
 where $R_{\rm A}$ is the standard \alfven\ radius.   Here, $m_1$ is the
mass of the neutron star in $M_{\sun}$, $R_6$ is its radius in $10^6$
cm, $L_{37}$ is the X-ray luminosity in $10^{37}$ \ergs, and $B_{12}$
is the neutron star surface magnetic field in $10^{12}$ G.  Assuming $m_1 = 1.25$, $R_6 \sim
1$, $L_{37}=30$, and $B_{12} \sim 10$, we obtain $R_{m} \sim 2\times10^8$ cm and $R_{\rm BB}/R_{m}\sim 0.7$.

We note that if we evaluate $L_{X}$ only in the range 0.5--10 keV, we
obtain $L_{X}\sim 1\times 10^{38}$ \ergs\ and $R_{\rm BB}\sim 9\times
10^{7}\ \rm cm$, similar to results obtained by ASCA in the same
energy range \citep{paul02}; this highlights the need to consider
higher energies where the bulk of the hard component is emitted.

\subsection{Timing Analysis}
\label{timinganalysis}

Our pulse-phase-averaged spectral analysis suggests that the soft and
hard spectral components for LMC X-4 have different physical origins
\citep[as discussed in detail by][]{hick04}.  We examine the time
variation of these components separately, by deriving energy-resolved
pulse profiles at the different superorbital phases.  We derive pulse
periods from the PIN observations, which have time resolution of 61
$\mu$s.  We are also able to study pulsations with XIS, which in 1/8
Window mode has time resolution (1 s) that is sufficient to produce
pulse profiles for LMC X-4's 13.5 s period (the XIS event arrival
times were randomized within the 1 s time bins).  Any variations in
source flux on timescales longer than the pulse period should not
significantly affect the derived pulse profiles.  Therefore, to
maximize signal to noise in the timing analysis, we used the full clean
GTIs for each detector (unlike for the spectral analysis, for which we
filtered the XIS and PIN data with a common GTI).

We corrected both the PIN and XIS events to the rest frame of the
solar system using the {\tt aebarycen} tool and corrected for the
orbital motion of the LMC X-4 system using the ephemeris of
\citet{levi00}.  Using the FTOOL {\tt efsearch}, we derived pulse
periods for the PIN data by searching for the maximum in the $\chi^2$
versus folding period.  We obtain $P_{\rm pulse} =
13.5087\pm0.0002$ s, $13.5091\pm0.0001$ s, and $13.5096\pm0.0001$ s
for observations 1--3, respectively.  Uncertainties are
estimated both from formal errors on the fit to the $\chi^2$ versus
period and also by a jackknife resampling technique, calculating the
period after excluding successive $\sim$4 ks GTIs;
these two estimates give consistent uncertainties. The drift in
$P_{\rm pulse}$ over the three months between observations 1 and 3 is
consistent with typical rates of spin evolution observed previously
for LMC X-4 \citep[e.g.,][]{woo96,levi00}.

As a check, we derived pulse periods from the XIS data, using the
combined event file for the XIS0, XIS1, and XIS3.  We obtain similar
pulse periods of $P_{\rm pulse} = 13.5089\pm0.0001$ s,
$13.5095\pm0.0001$ s, and $13.5100\pm0.0003$ s; the resulting pulse
profiles are qualitatively identical if we use these periods rather
than those from the PIN.

We note that we have not accounted for any deviation from orbital
ephemeris derived by \citet{levi00}, which could possibly affect the
derived spin periods.  However, the relative pulse profiles in the
different X-ray bands, which are the focus of this study, will be
largely independent of the exact absolute spin period.  We leave a
detailed analysis of the spin evolution of LMC X-4 to a future study.

Using the above spin periods, we created separate pulse profiles for
the XIS in the 0.5--1 keV (``soft'') band (which is dominated by the
soft blackbody spectral component) and the 2--10 keV (``medium'')
band (which is dominated by the hard power-law component).  We also
derived profiles for the PIN in the 10--50 keV (``hard'') band that is
entirely dominated by the hard (cutoff) power law.  We have subtracted
the background flux in producing the PIN pulse profiles (the XIS
background is negligible).

\begin{figure}
\plotone{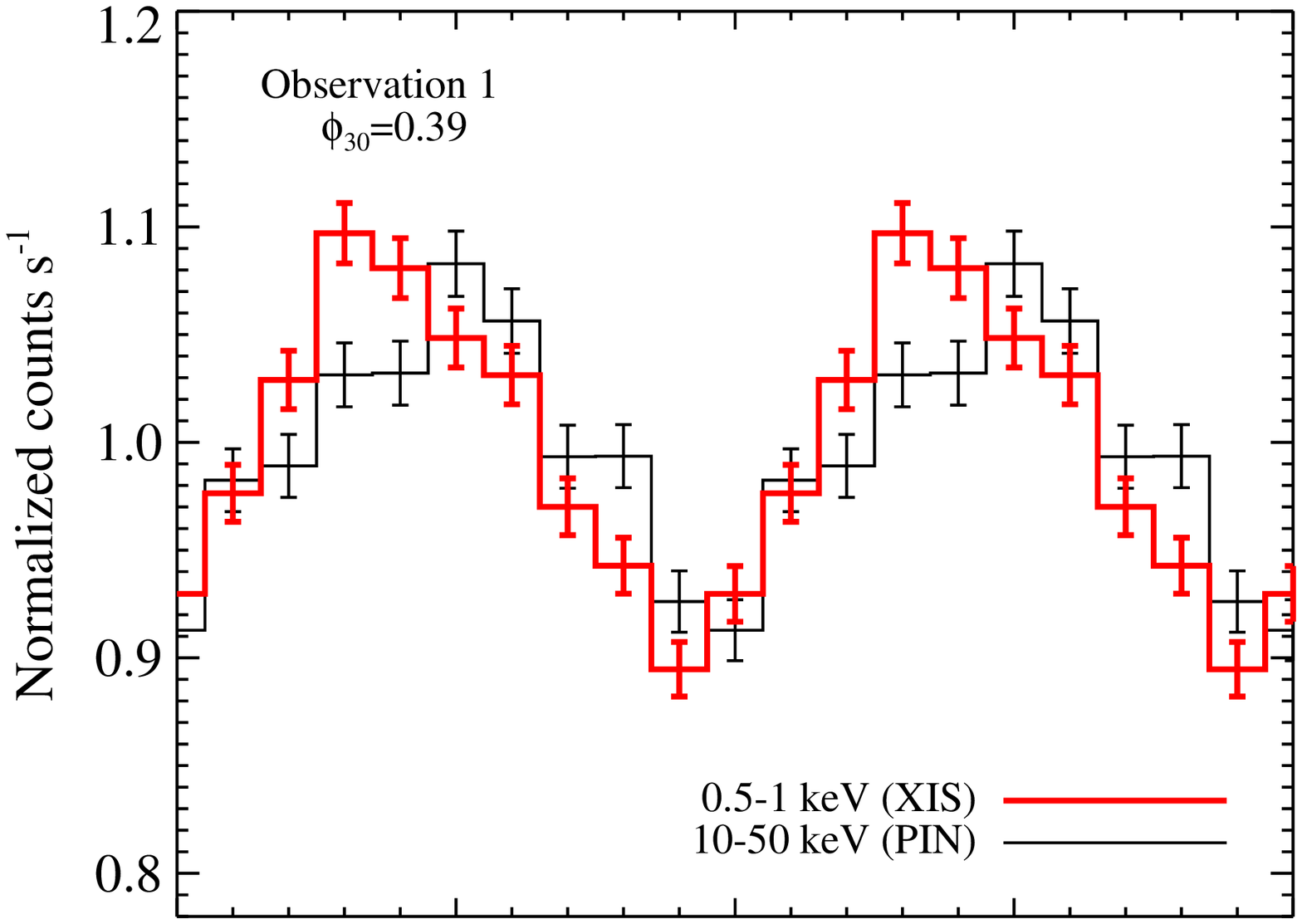}
\vskip-1.7cm
\plotone{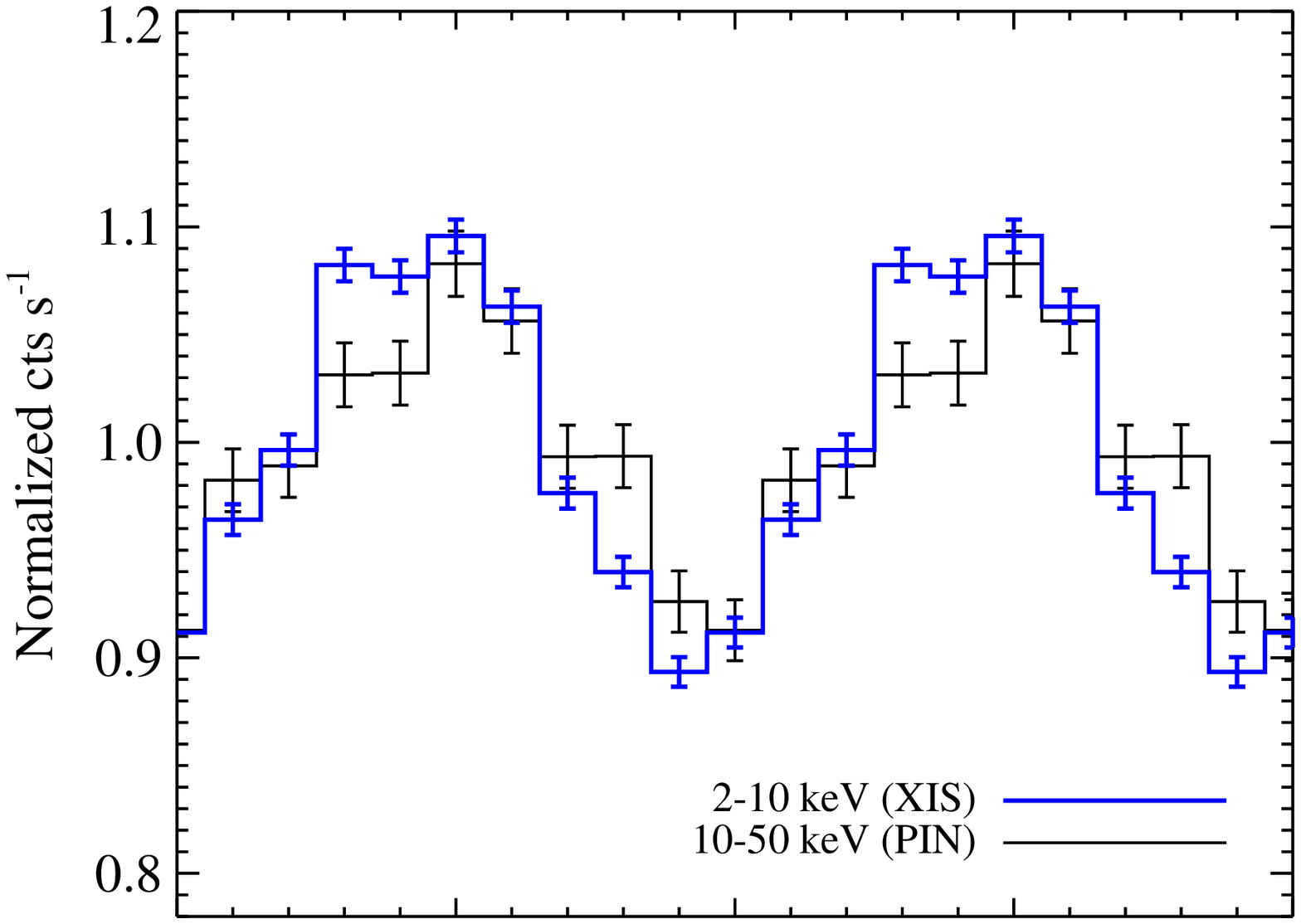}
\vskip-1.7cmi
\plotone{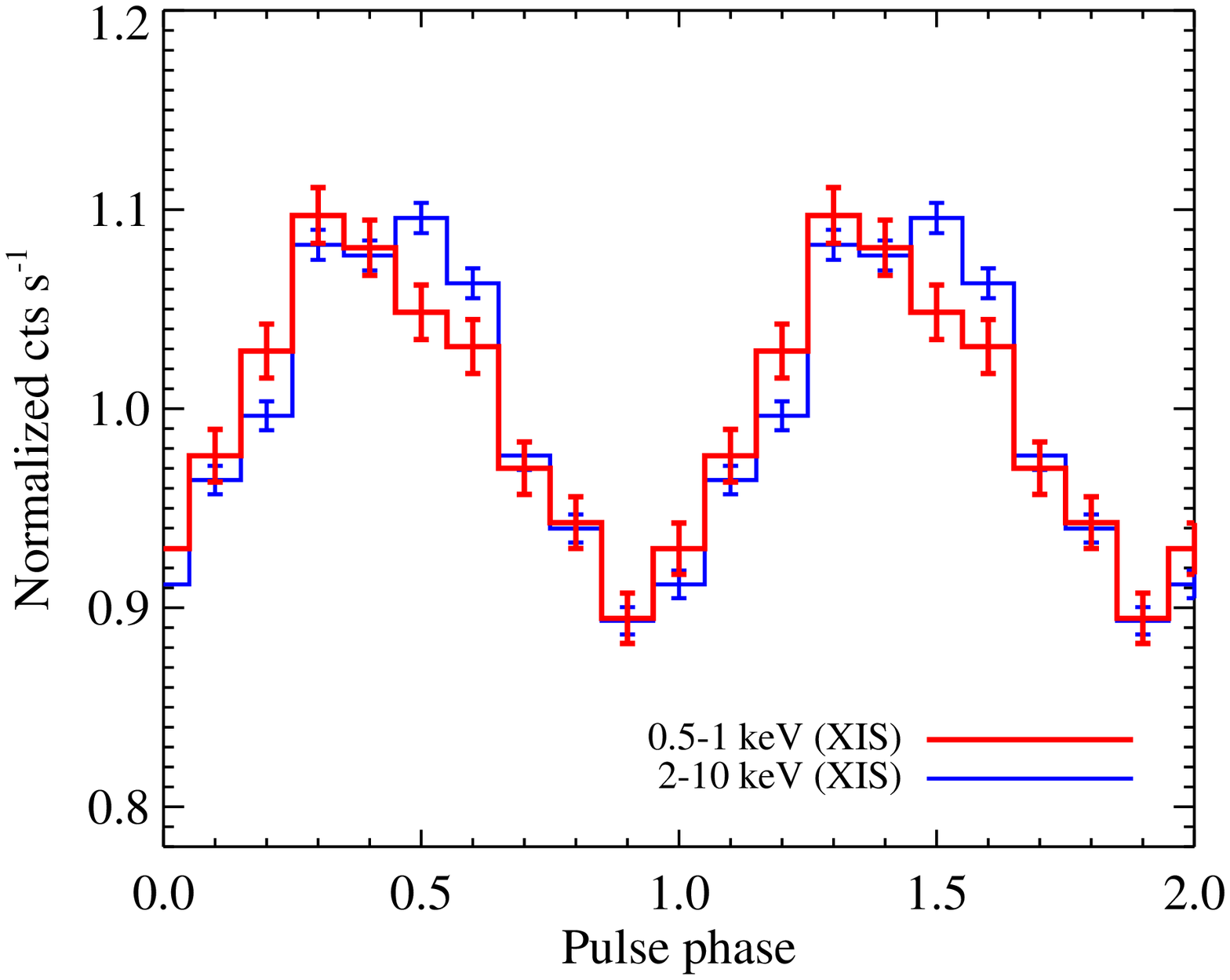}
\caption{Energy-resolved pulse profiles for observation 1.  The soft
  (0.5--1 keV, red) pulses are dominated by the blackbody component,
  while the medium (2--10 keV, blue) and hard (10--50 keV, black)
  pulses are dominated by the power law.  The profiles for
  observations 2 and 3 are shown in Figures~\ref{pp2} and \ref{pp3},
  respectively; note the variation in the medium pulse profile and the
  varying phase offset between the hard, medium, and soft pulses for
  the different superorbital phases. \label{pp1}}
\end{figure}

\begin{figure}
\plotone{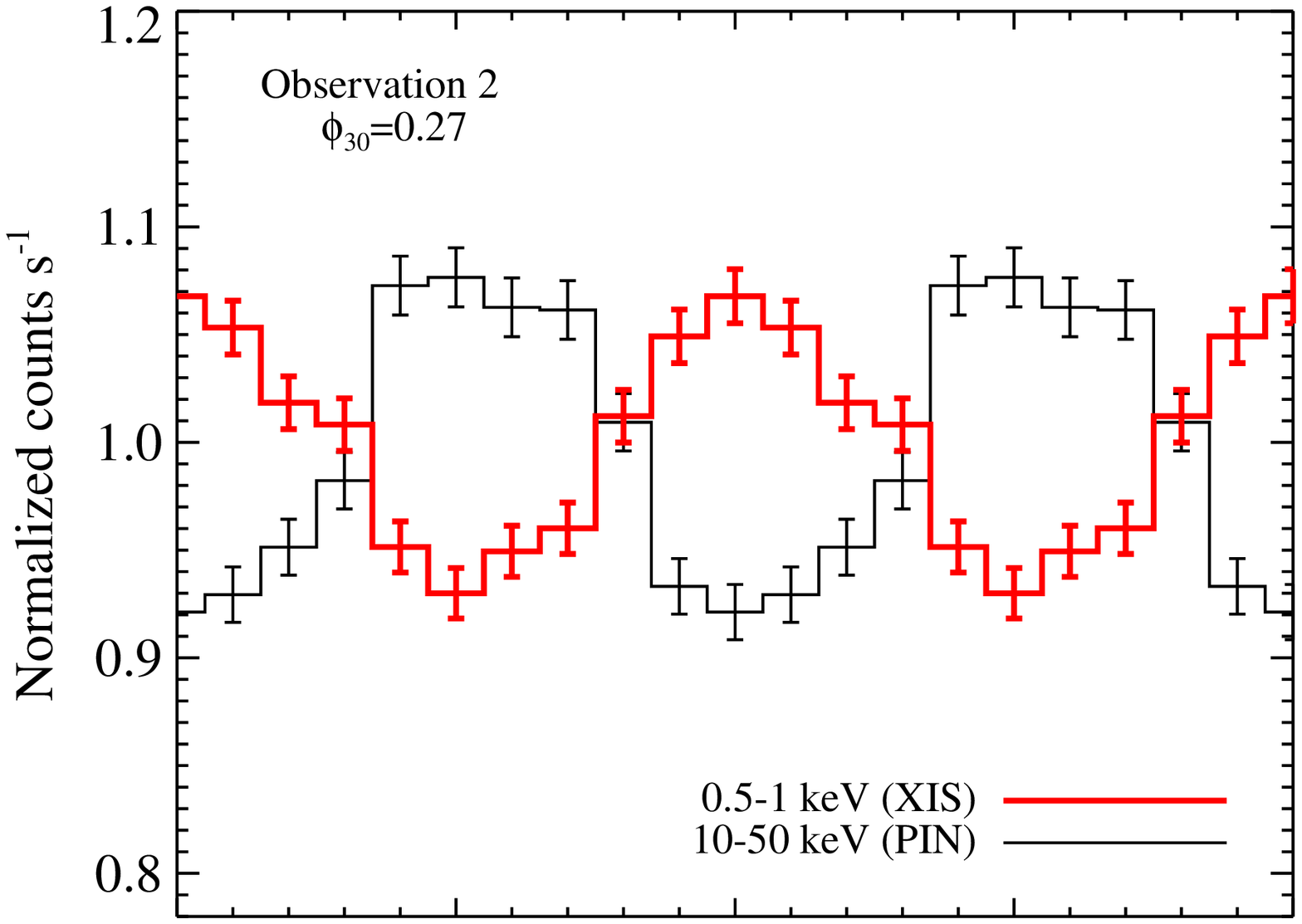}
\vskip-1.7cm
\plotone{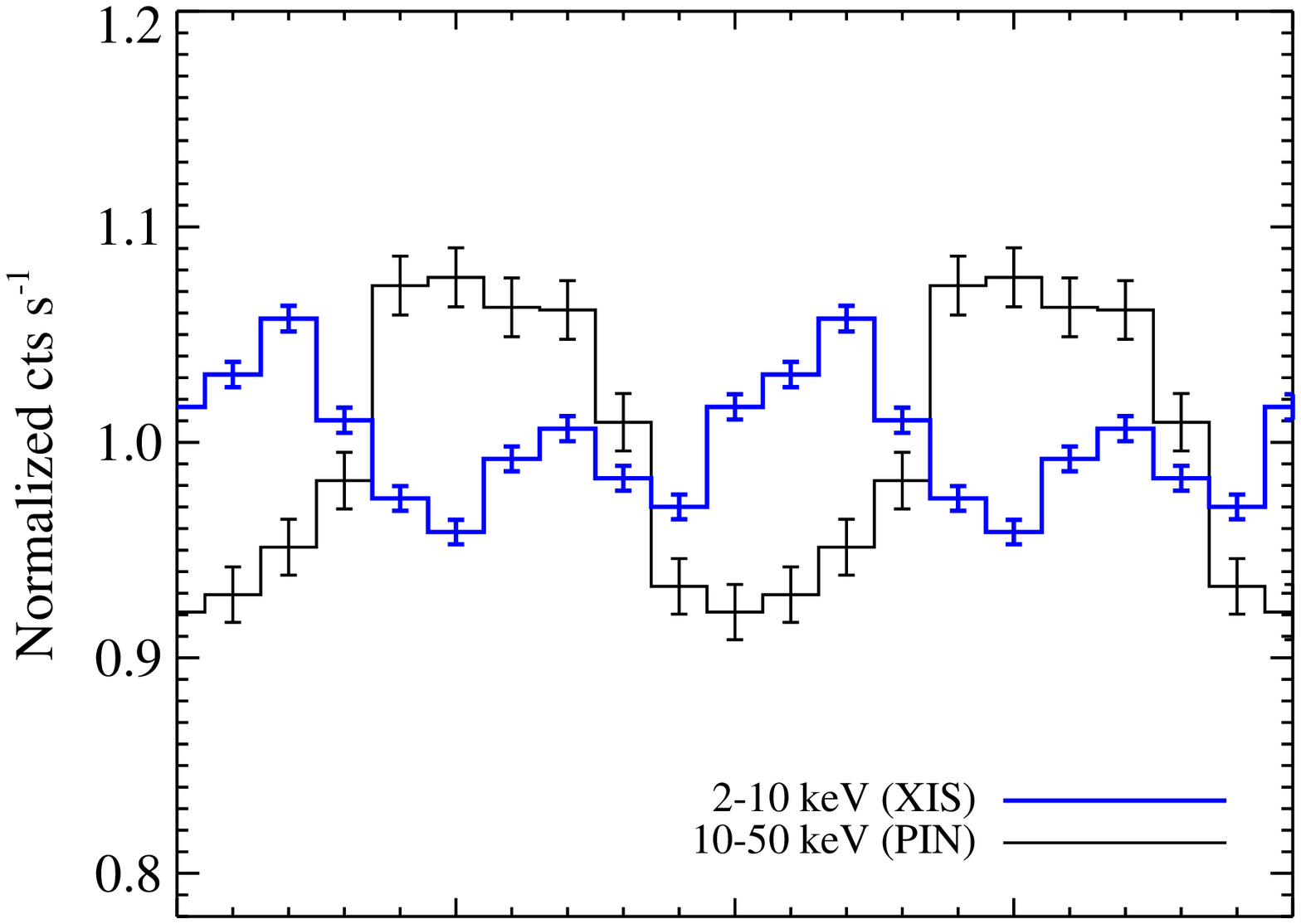}
\vskip-1.7cm
\plotone{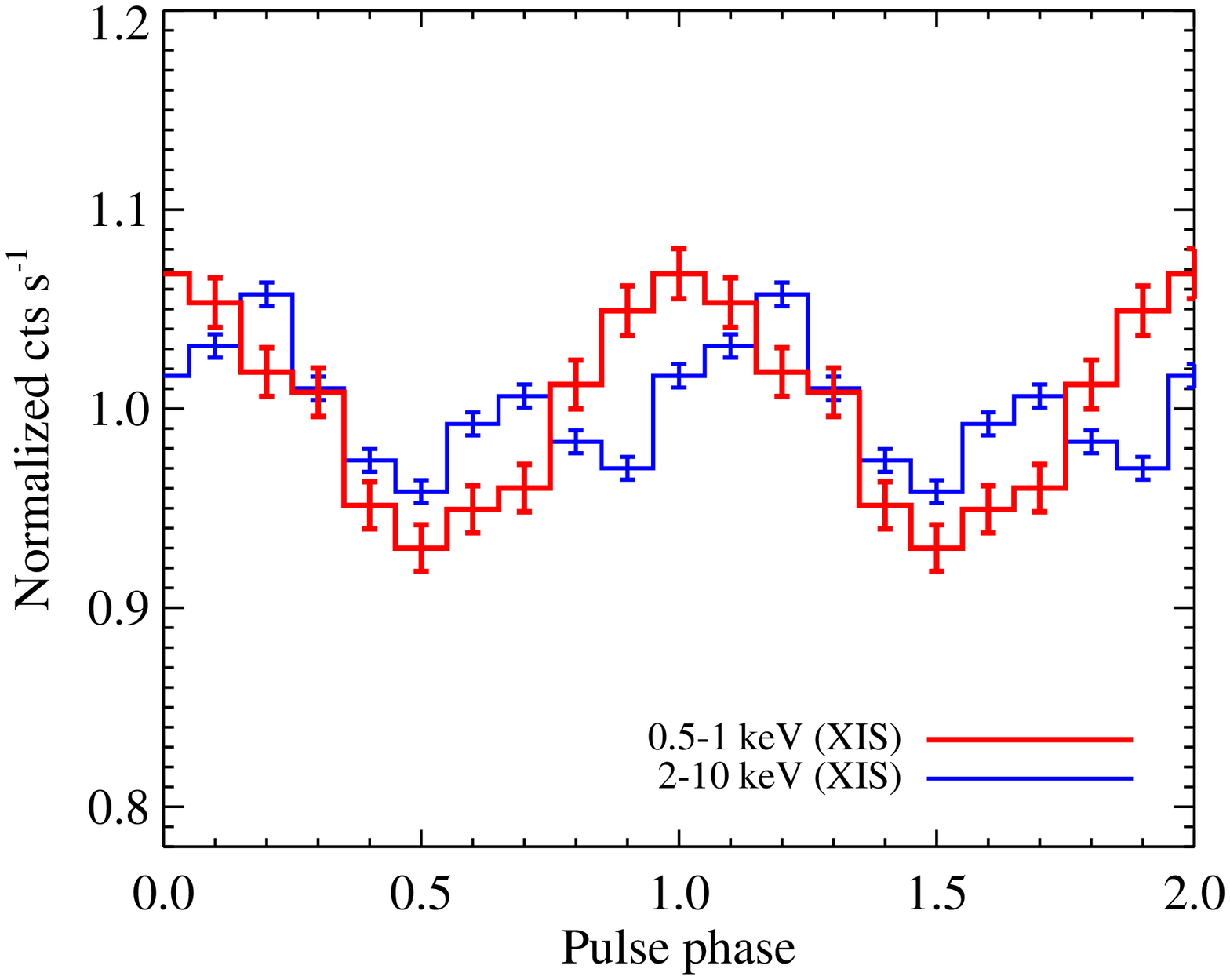}
\caption{Energy-resolved pulse profiles (as in Figure~\ref{pp1}), for observation 2.  \label{pp2}}
\end{figure}

\begin{figure}
\plotone{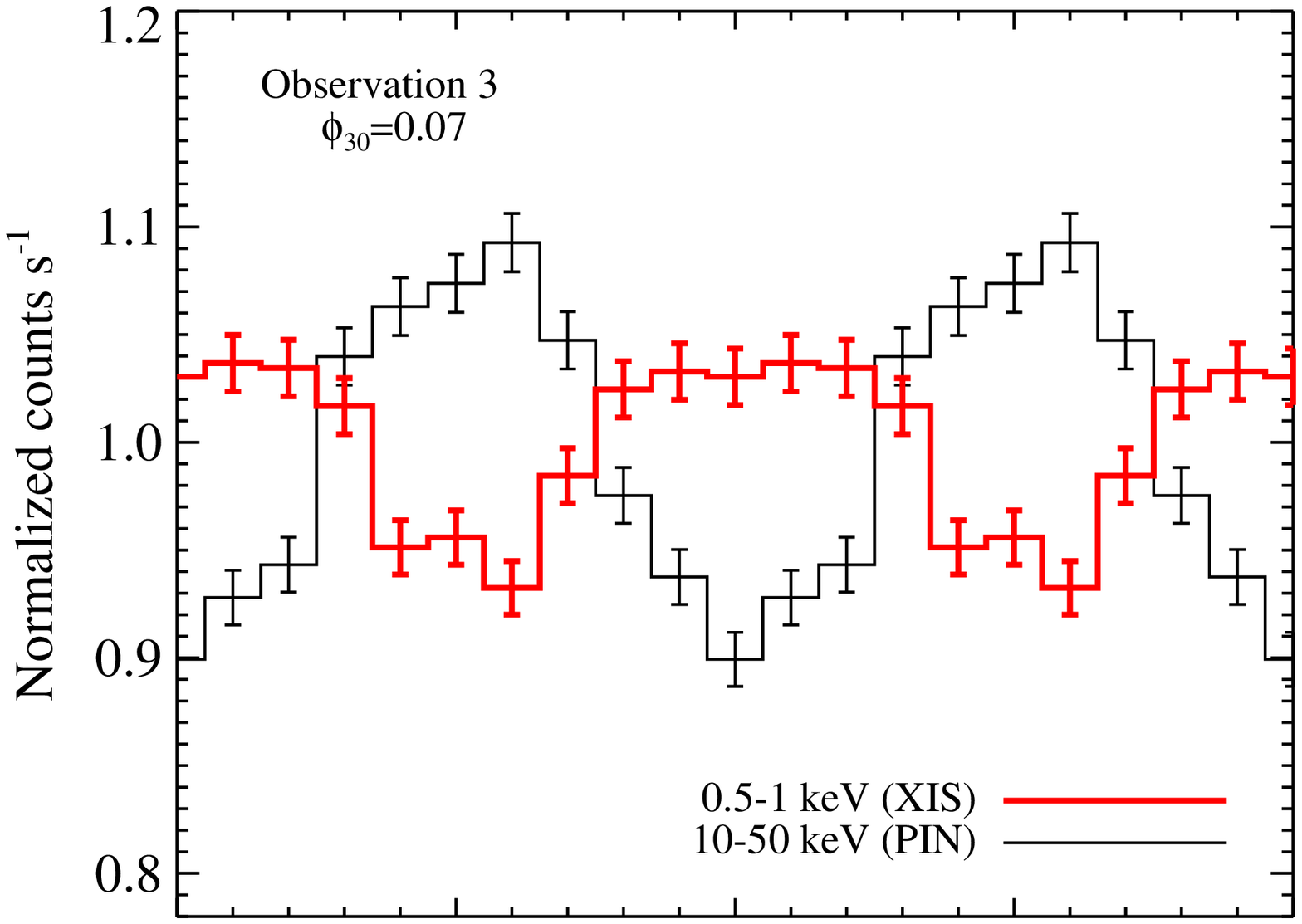}
\vskip-1.7cm
\plotone{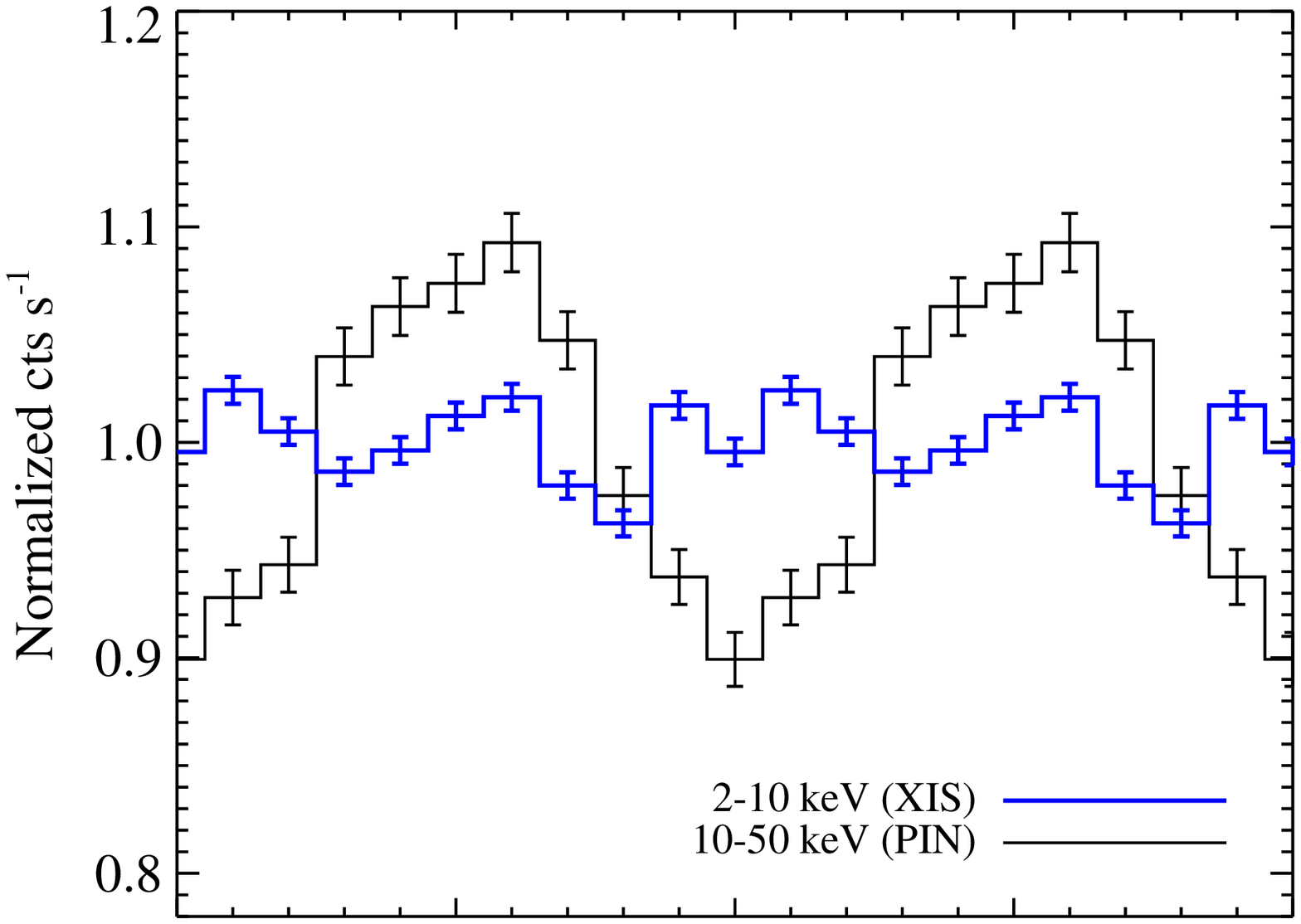}
\vskip-1.7cm
\plotone{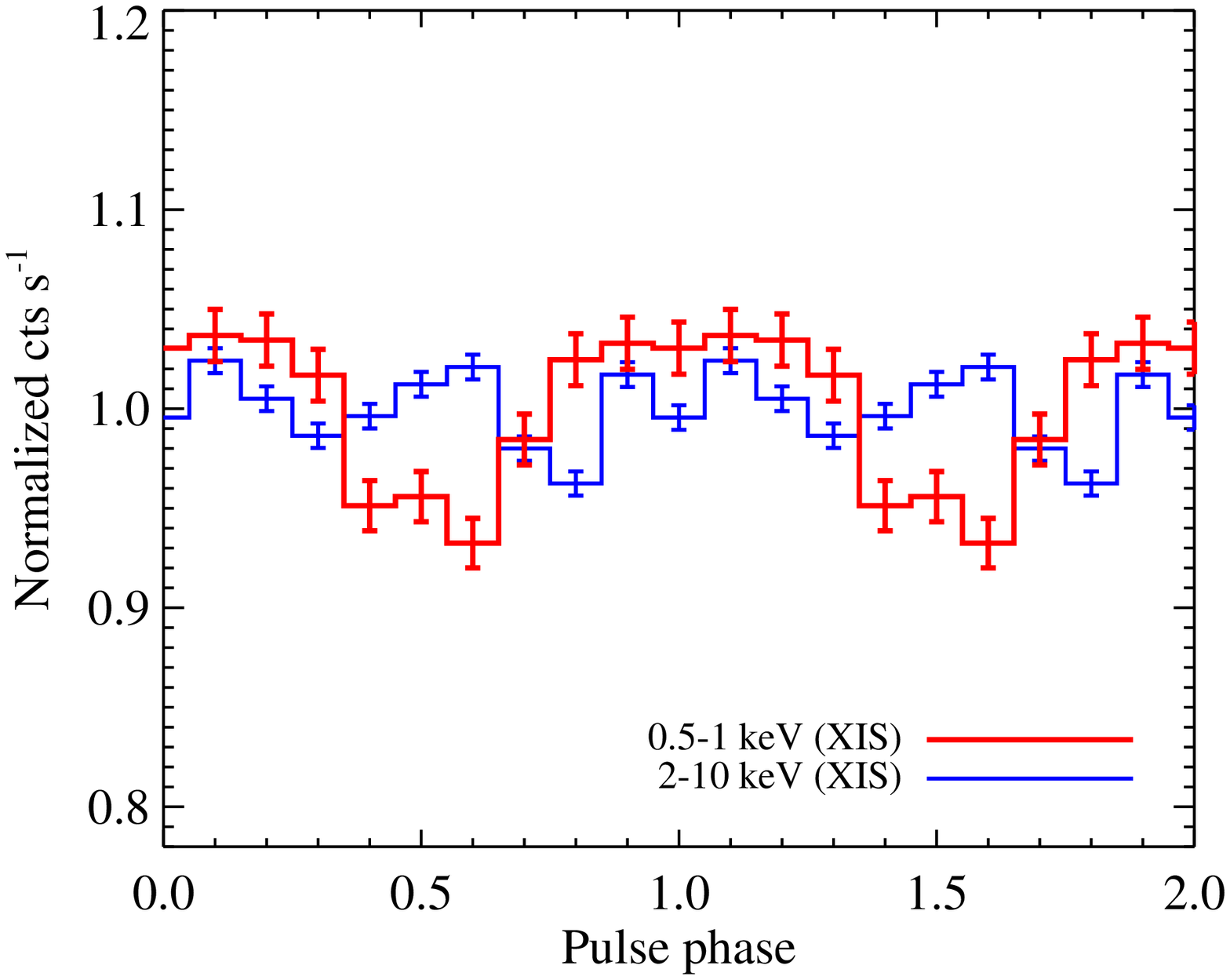}
\caption{Energy-resolved pulse profiles (as in Figure~\ref{pp1}), for observation 3. \label{pp3}}
\end{figure}

The pulse profiles are shown in Figures~\ref{pp1}--\ref{pp3} and show
significant differences in the pulse profile for the different X-ray
energies.  The soft and hard pulses in all three observations are
broad and show single peaks, while observations 2 and 3 show
double-peaked profiles in the medium band, similar to those observed
in Her X-1 \citep{zane04} and SMC X-1 \citep{neil04,hick05}, although
the pulsations for observation 3 are relatively weak.  In contrast
observation 1 shows only a single peak in the medium pulse profile,
similar to the observed soft pulses.  The variations between the hard
and medium pulses indicate that the spectral shape of the medium
component may change with pulse phase and further that these
variations may differ between observations.

\epsscale{1.2}

\epsscale{1.2}
\begin{figure}
\plotone{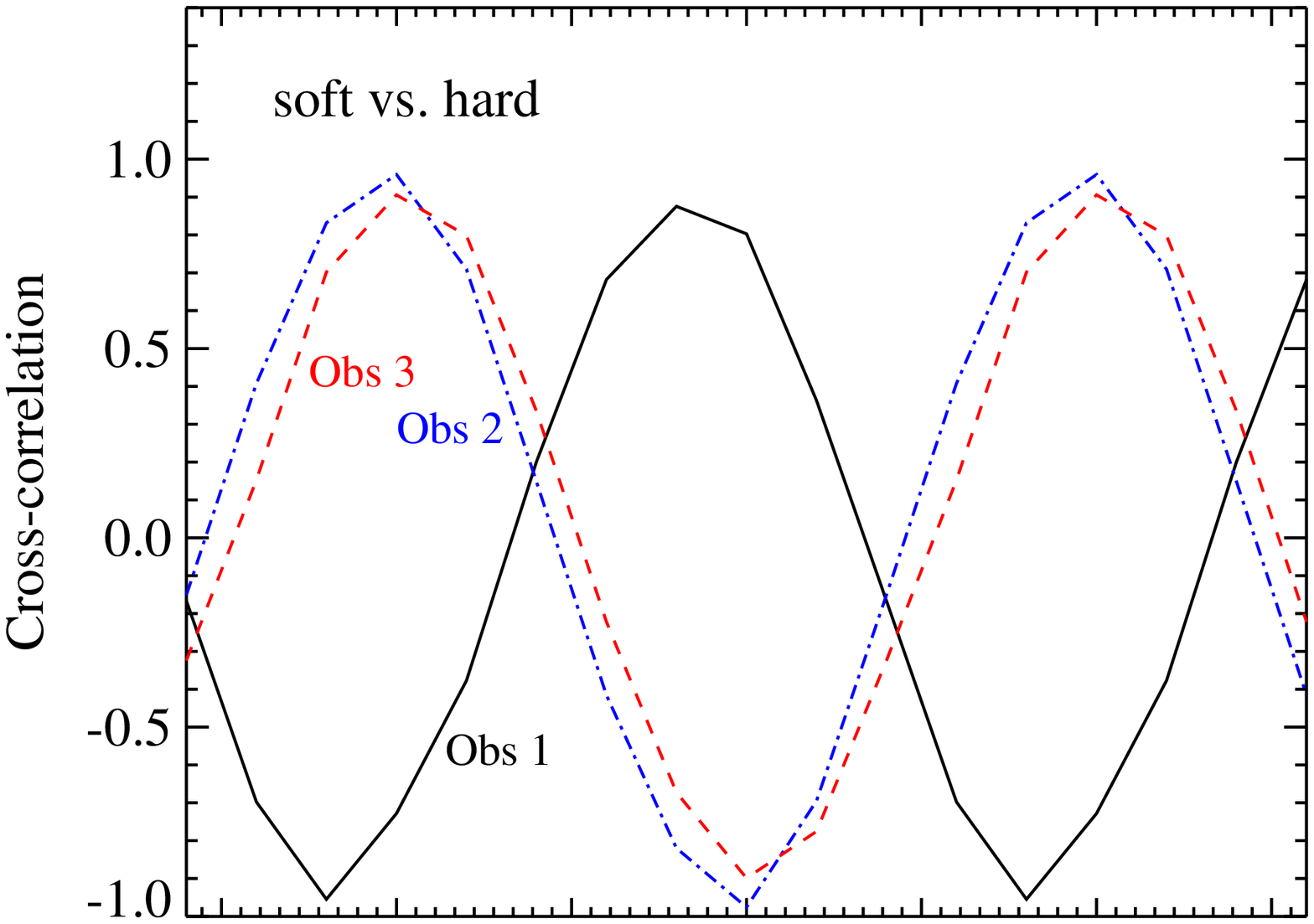}
\vskip-1.7cm
\plotone{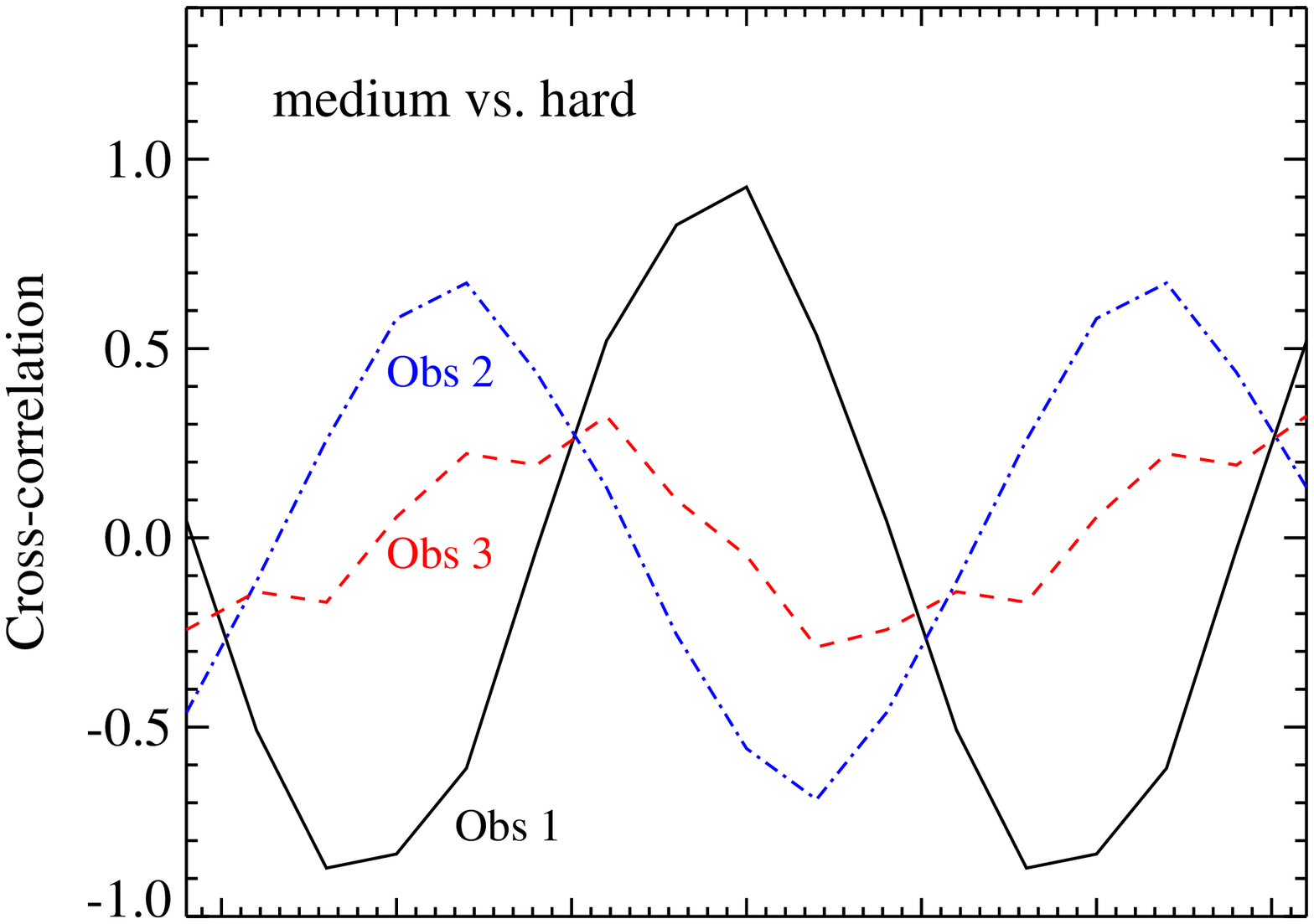}
\vskip-1.7cm
\plotone{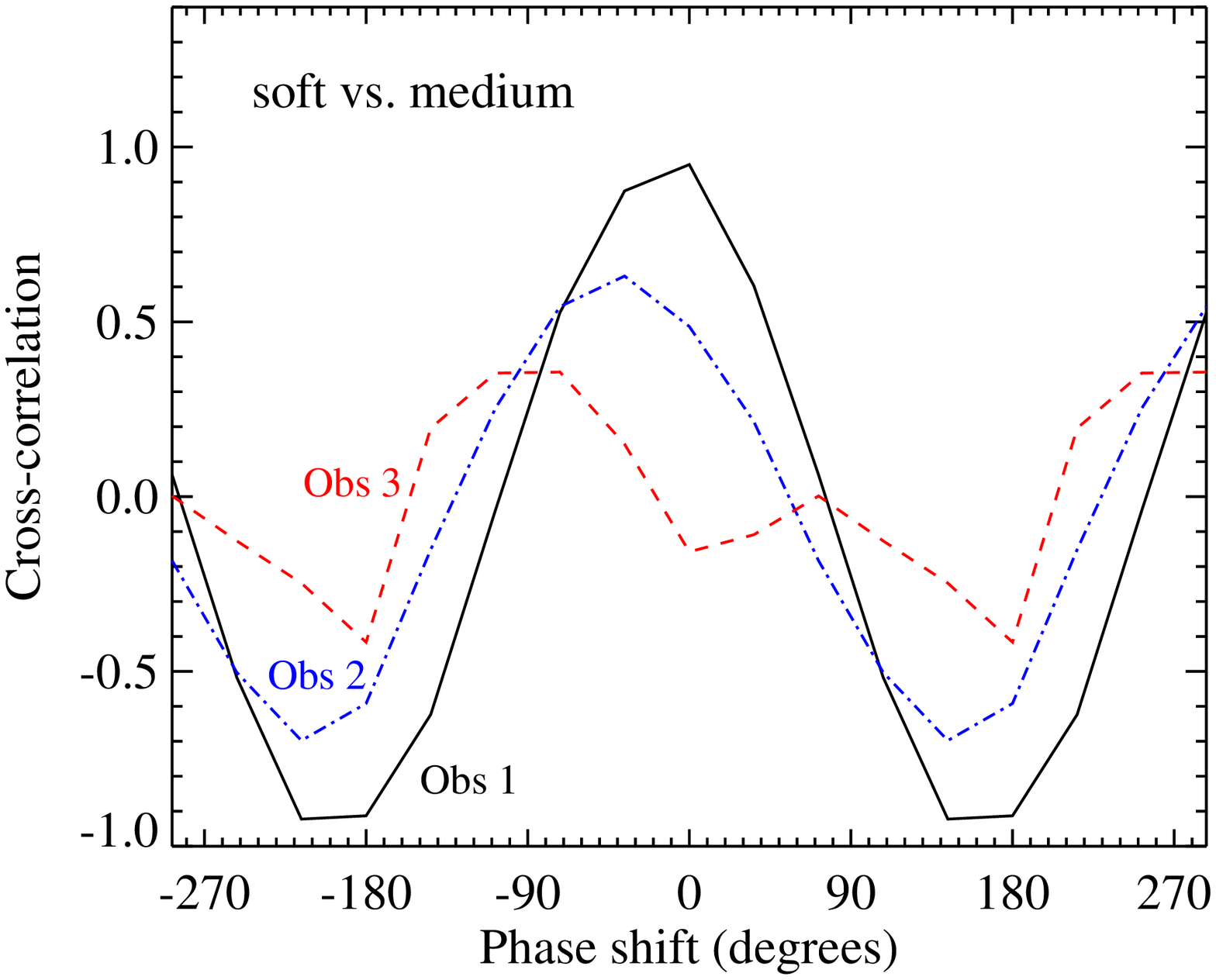}
\caption{Cross-correlation between energy-resolved pulse profiles,
  showing the variation in relative phase and shape of the pulses for
  the three \suzaku\ observations.  A positive phase shift corresponds
  to the first pulses listed leading the second pulses listed.
  For example, in the top panel, the peak at $\sim-40$\dgr\ shift for
  observation 1 indicates that the soft pulses lag the hard pulses by
  $\sim$40\dgr.
\label{cc}}
\end{figure}

\epsscale{0.8}
\begin{figure}
\plotone{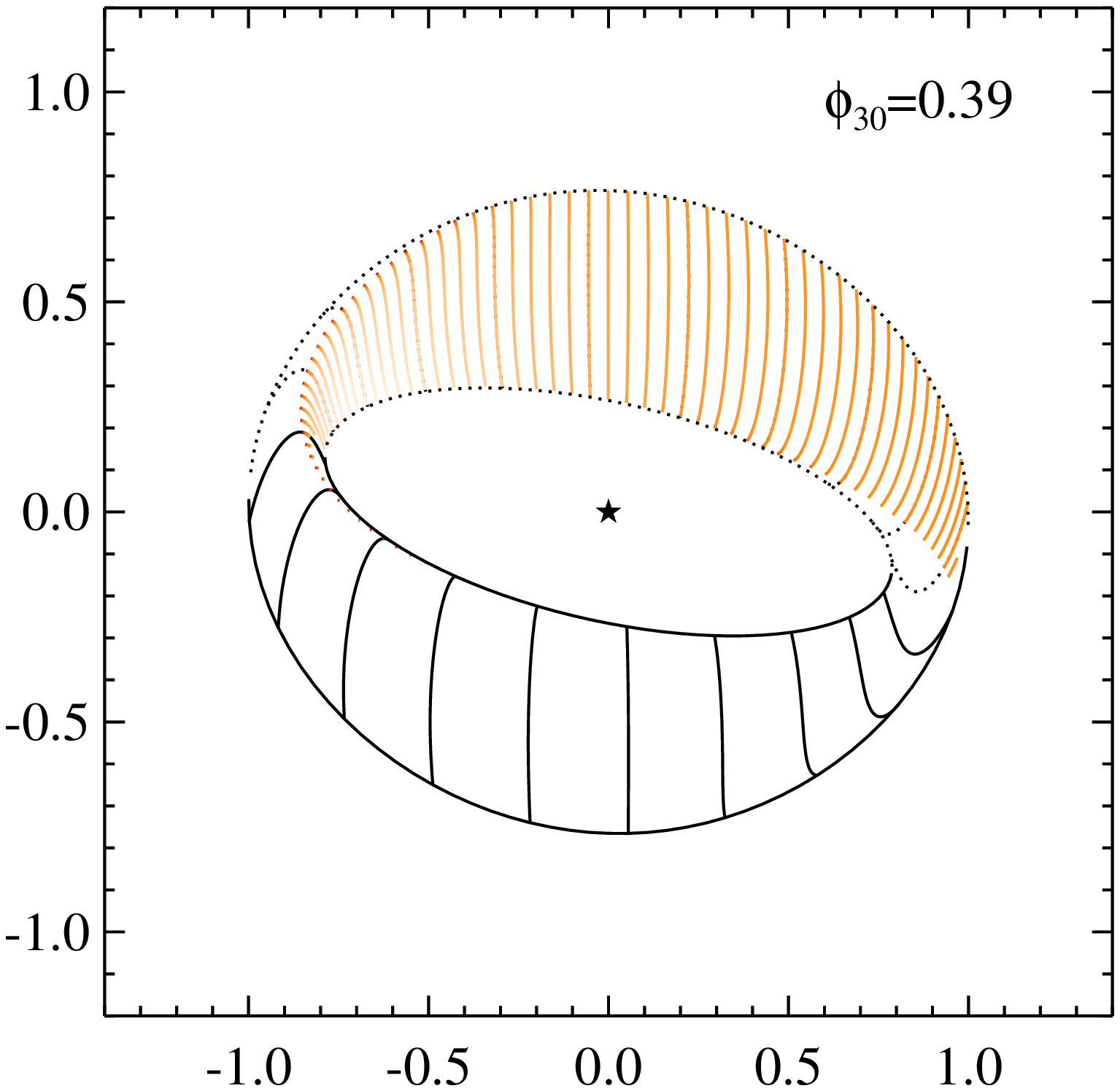}\\
\plotone{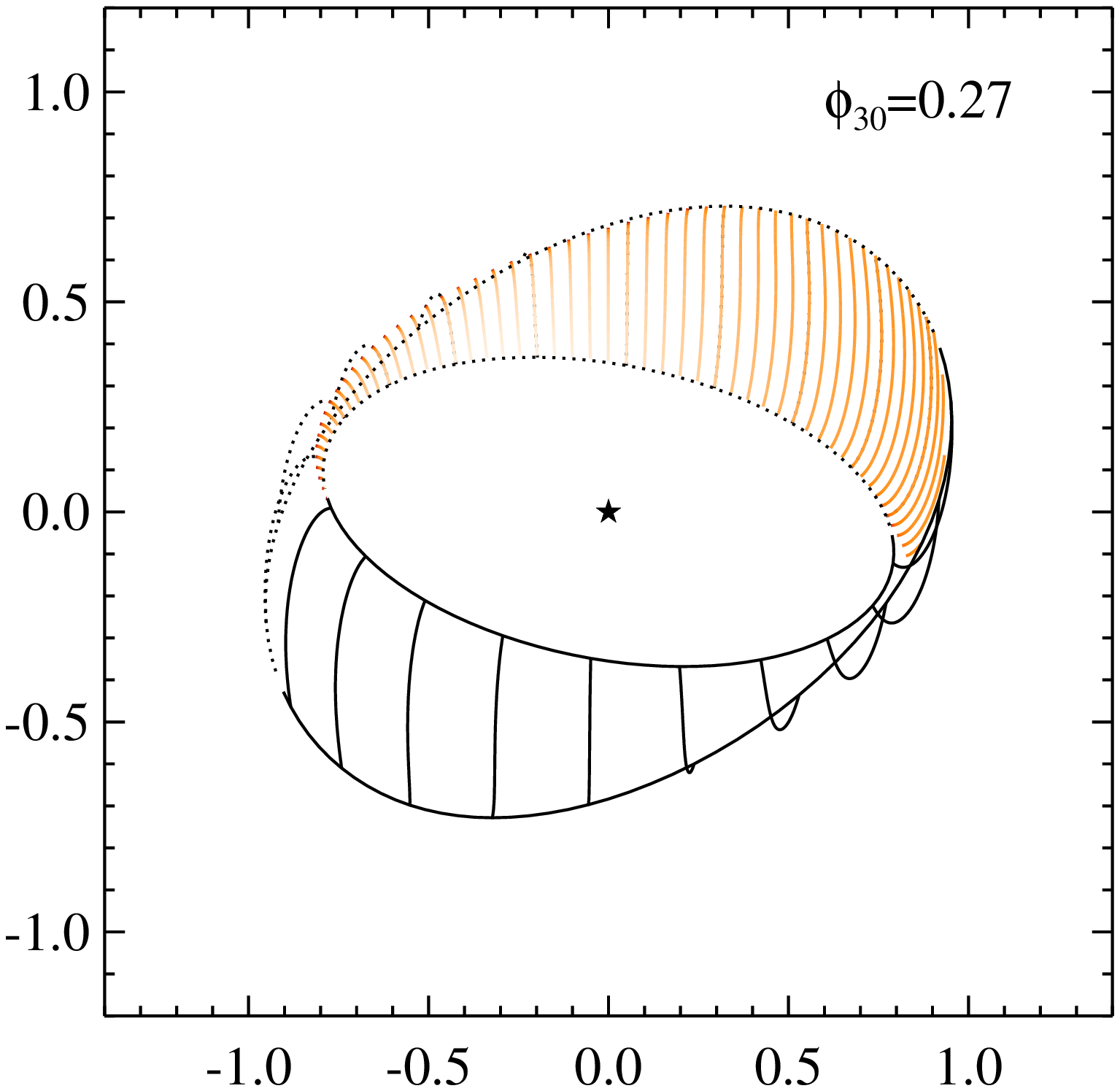}\\
\plotone{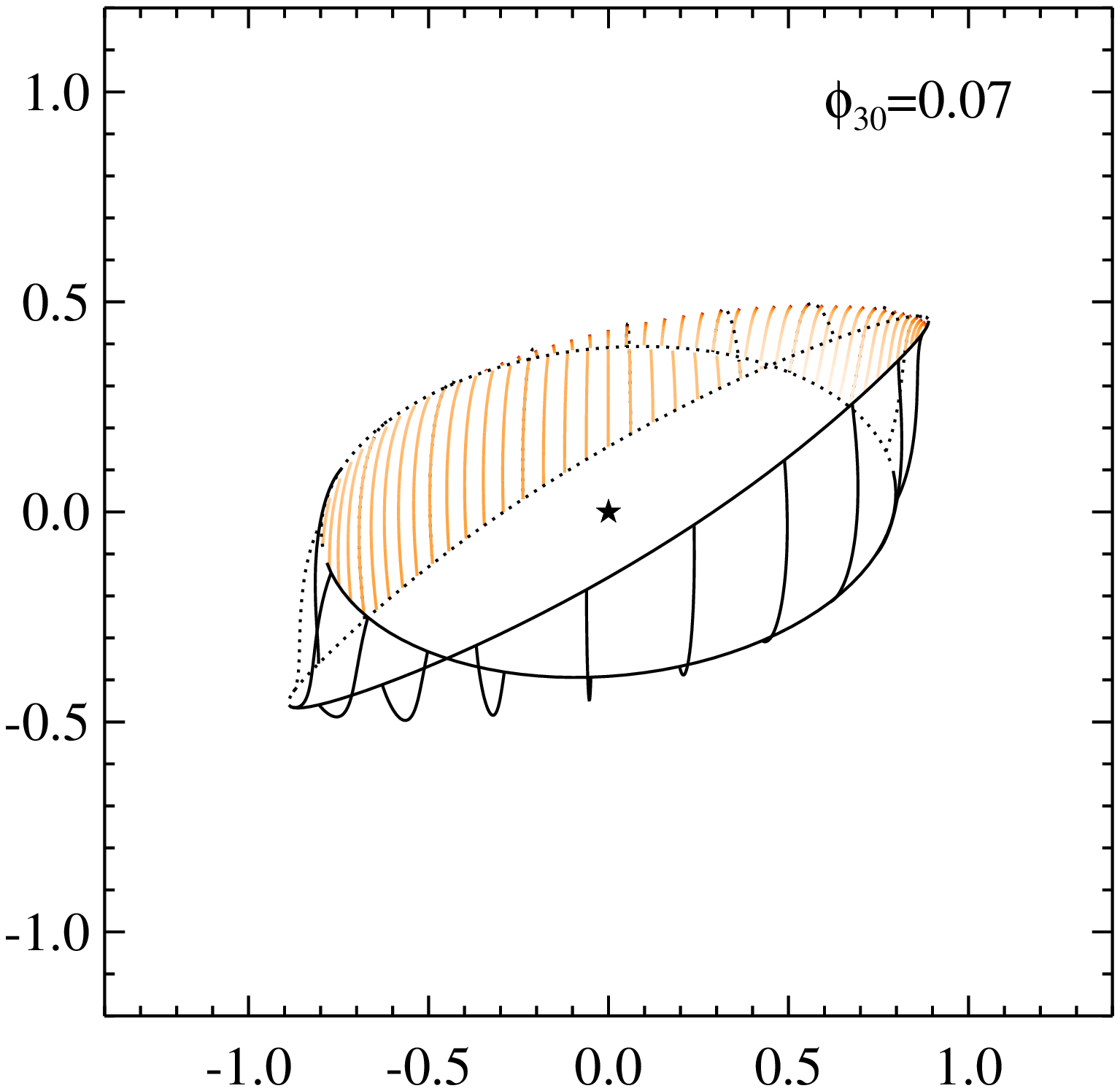}
\caption{Model of the inner regions of a twisted, warped accretion
  disk, illuminated by the neutron star and observed at three
  superorbital phases corresponding to the {\em Suzaku} observations.  
  We set the inclination of the disk to be 70\dgr\ \citep{meer07}.
  The orange shaded regions show areas on the disk which are
  illuminated by the central star and are visible to the observer.  \label{model}}
\end{figure}

In addition, the phase shifts between pulses at different energies
vary among the observations. In observation 1, all three pulses are
almost in phase with each other, while observations 2 and 3 show
notable phase shift.  We study this phase shift in more detail by
cross-correlating the soft profiles against the hard and medium
profiles (see Figure~\ref{cc}).  We first examine the cross-correlation
between soft (0.5--1 keV) and hard (10--50 keV) pulses, shown in the
top panel of Figure~\ref{cc}.  These energy bands represent the bulk of
the blackbody and hard power flux, respectively.  Observation 1 shows
a strong positive cross-correlation between soft and hard pulses
peaking at $\sim$-40\dgr, indicating that the soft pulses slightly lag
the hard pulses.  The cross-correlations for observations 2 and 3 peak
almost exactly out phase at $\sim$180\dgr, although there is an
indication that the observation 3 may be slightly shifted to shorter
phase lag (by $\sim$10\dgr). However, the precise relative lag is
difficult to measure because the pulse profiles have relatively few
(10) bins.

The cross-correlations between the medium (2--10 keV) pulses and the
hard and soft pulses are shown in the bottom two panels of
Figure~\ref{cc}.  For observation 1, the cross-correlations peak near
zero as the phase of the medium pulses closely aligned with the soft
and hard pulses.  The cross-correlations for observations 2 and 3 are
more complex, because of the multiple peaks in the medium pulse
profiles.  We note that the phase offset between medium and hard pulses varies more
strongly than between medium and soft pulses.


\section{Discussion}
\label{discussion}

We use 13 years of available \rxte/ASM data to determine a
superorbital period of 30.32$\pm$0.04 days for LMC X-4.  The long time
series allows an improvement by a factor of 10 in precision compared
to the measurement of \citet{lang81}.  Our period estimate is
consistent with that of \citet{lang81}, indicating that the period has
been relatively stable over the past $\sim$30 years.  The improved
period allows us to accurately determine the superorbital phases of
our \suzaku\ observations and will facilitate future studies that
require accurate estimates of $\phi_{30}$.

In the \suzaku\ observations, we find that the continuum over roughly
two decades in energy (0.6--50 keV) can be described by a blackbody
plus a cutoff high-energy power law.  In addition, we detect four
emission features as listed in Table\ \ref{lines}.  (All of the four emission lines have been detected in
SMC X-1; \citep{vrti05}.)  Limited by the energy resolution ($\sim130$ eV) of XIS detectors, we
cannot clearly identify the line with energy 0.98 keV.  Our possible
candidates are Ne\,{\sc ix} (0.91 keV) and Ne\,{\sc x} Ly$_\alpha$ (1.02 keV).  For
the Fe K$_\alpha$ line, it has been detected from many previous
studies \citep[e.g.,][and references therein]{laba01,naik03}.
\citet{neil09} suggested that Fe K$_\alpha$ line originates at the
outer accretion disk.  In addition to Fe K$_\alpha$ at 6.4 keV, we
also detect a broad Fe line.  Using $M$ = 1.25$\msun$ for the neutron
star and $R_{\rm BB}\sim 1.5\times 10^{8}\ \rm cm$, we find the
velocity at the inner accretion disk to be $\sim 1.1\times 10^{4}$
km s$^{-1}$, which corresponds to Doppler shift of 0.2 keV for Fe K$_\alpha$.
This is of similar order (although somewhat smaller than) the average
$\sigma\ \approx 0.4$ keV derived from the spectral fits.  Given this
rough correspondence (and the considerable uncertainties in the
geometry of the accretion flow), we argue that this broad Fe line is
Doppler broadened Fe K$_\alpha$ emitted from the inner accretion disk
due to the high velocity, as also suggested by \citet{neil09}.

It is challenging to estimate the fluorescent Fe line flux that would
be expected for reprocessing by the inner disk.  For the partial
spherical geometry described above with $\Omega = 0.05$, a model of
reflection from a neutral disk \citep{geor91reflect} predicts an Fe K
line ${\rm EW} \sim 25$ eV, similar to the EW of the observed narrow line
but smaller than the broad Fe feature, which has ${\rm EW} = 150\pm50$ eV.
 We note, however, that such a broad feature could be produced by
emission from iron in a range of ionization states \citep[e.g.,][]{rams02} and
could also be due to complex absorption around the Fe K edge, as
observed in some active galactic nuclei \citep[e.g.,][]{yaqo95ngc4151}
and polars \citep[e.g.,][]{done98bycam}.  Finally, disk reflection would
predict the presence of a Compton reflection feature which we have not
included in our simple power-law model for the continuum, and which
might affect the observed EW.  In light of these uncertainties, it is notable that a broad Fe feature
with similar ${\rm EW}\sim 150$ eV has also been observed in Her X-1, for
which correlations in the pulse profiles for the soft (blackbody)
component and the Fe K line suggest that both originate from the inner
disk \citep{rams02,zane04}.  We conclude that inner disk reprocessing
is a plausible explanation for the observed Fe line emission, although
further work is needed to robustly determine the physical origin of
the broad feature.

The pulse profiles of LMC X-4 show that the profiles at different
energies have different pulse behaviors.  In all three
observations, the soft (0.5--1 keV) and hard (10--50 keV) profiles
show single-peaked, smooth, roughly sinusoidal pulsations.
Single-peaked soft pulses have been observed in the well-studied X-ray
binary pulsar systems, Her X-1 and SMC X-1 \citep{zane04,hick05}.
\citet{hick05} using a simple pencil beam model with a twisted inner
accretion disk were able to reproduce the observed single-peaked
sinusoidal pulsations.  We suggest that the soft pulses of LMC X-4 are
also due to reprocessing of hard X-rays in the inner accretion disk.

The medium (2--10 keV) component, however, shows single-peaked pulses
in observation 1 and double-peaked pulses in observation 2 and
(although weakly) in observation 3.  Multiple-peaked  pulses at $E>2$ keV have
been reported several times in the literature; \citet{woo96} observed
double-peaked pulses with one dominant peak, while \citet{naik04}
and \citet{paul02} find pulses with three equally strong peaks.
Weak or absent pulsations have also been observed \citep{laba01}.

Further, the phase shift between the pulsations changes significantly
between the three observations.  In particular, in observation 1 the
pulses in all three bands are largely in phase, while for observations
2 and 3, the hard and soft pulses are almost 180\dgr\ out of phase,
while the medium pulses are shifted somewhat less relative to the soft
pulses.

Following previous studies of Her X-1 \citep[e.g.,][]{rams02,zane04}
and SMC X-1 \citep{hick05}, we attempt to interpret the observed pulse
profiles in terms of the soft component being reprocessed emission
from the inner disk.  We begin by considering a fiducial model for
the inner regions of the accretion disk at the superorbital phases of
the three observations (Figure~\ref{model}).  The model of a warped,
twisted disk is the same as that used by \citet{hick05} to
describe the superorbital variation in SMC
X-1\footnotemark. \footnotetext{The disk corresponds to the best-fit
  model for a pencil beam, with parameters given in Table 2 of
  \citet{hick05}.}  We assume a disk inclination of 70\dgr, consistent
with the orbital inclination of LMC X-4 \citep{meer07}.  The area of
the disk illuminated by the neutron star is shown in the shaded
region.  While we do not have strong constraints on the shape of the
inner disk for LMC X-4, this model gives us a qualitative picture of
what we might expect for reprocessing by a warped, precessing disk.

Another consideration is the geometry of the pulsar beam.  For SMC
X-1, \citet{hick05} found that pencil- or fan-shaped beams from the
neutron star can provide an acceptable description of the observed
pulse profiles, although the pencil beam is preferred at some
superorbital phases.  For the super-Eddington luminosities observed
for SMC X-1 and LMC X-4, theoretical models generally predict that
radiation cannot escape from the top of the accretion column and so
favor a fan beam emitted from the walls of the column
\citep[e.g.][]{beck98,beck05}.  However, fan beam models universally
predict a double- or quadruple-peaked hard pulse profile, in contrast
to the single peaks observed in LMC X-4.  Further, fan beams can
occupy a wide and complex range of geometries (for example the reverse
fan beam suggested by \citealt*{scot00} for Her X-1), and a full
treatment requires a detailed description of the accretion flow as
well as effects such as gravitational light bending \citep[e.g.,][]{mesz88,
  leah03}. Given these difficulties, in the following discussion we
consider only a simple pencil beam and leave a precise determination
of the beam geometry for future studies.

Assuming that the radiation pattern consists of two pencil-shaped
beams pointing in opposite directions, then if the beams are not
exactly in the plane of the accretion disk, one beam will produce a
strong observed hard pulse (as we observe for LMC X-4) while the other
beam may more strongly illuminate the disk.  In observation 1, the
largest visible area of the disk is almost exactly in the
line of sight, and we may expect the hard and soft pulses to be
almost in phase with each other.  Figure~\ref{model} shows that as the
disk precesses, the largest visible area of the disk is shifted to the
side, which would suggest that the phase shift between the pulsations
from the neutron star and those from the reprocessing disk should vary
with superorbital phase.  This is indeed what we observe for the shift
between the medium and soft X-ray pulses (shown in the bottom panel of
Figure~\ref{cc}), for which the peak of the cross-correlation increases
monotonically between the three observations.  Such variations in the
$\sim$2--10 keV and $\sim$0.5--1 keV pulse profiles have been
interpreted similarly for Her X-1 \citep{zane04} and SMC X-1
\citep{neil04}.

However, these studies were limited to X-ray data below 10 keV.  With
the existence of the higher-energy PIN observations, it makes sense to
consider the hard (10--50 keV) emission to be the best indicator of
the luminosity in the neutron star beam, as the spectrum peaks at
these energies.  As discussed above, the hard and medium pulses show
significantly different behavior.  The cross-correlations in
Figure~\ref{cc} indicate that the hard pulses are either close to in
phase (observation 1) or close to completely out of phase
(observations 2 and 3).

This behavior suggests a different interpretation for the variation in
the phase offsets with $\phi_{30}$.  One idea is that if the beam
aligned with the observer also illuminates the largest visible
area of the accretion disk, then we might expect the hard and soft
pulses to be almost exactly out of phase.  It is therefore possible
that between observations 1 and 2, the hard X-ray beam and the visible
region of the disk moved abruptly relative to each other, thus
changing which of the two pulsar beams is illuminating the visible
region of the disk.  In terms of this picture, the observations would
be consistent with a simple ``donut'' geometry for the reprocessing
region, rather than the more complex twisted disk as shown in
Figure~\ref{model}.

We further note that the amplitudes of the soft pulses are equal to
or slightly weaker than those of the hard pulses.  If the pulses arise
from reprocessing, we might expect the reprocessing to blur out the
soft profile and thus decrease the amplitude relative to the hard
pulses.  The observed amplitudes might be explained either if the bulk
of the emission comes from a relatively small reprocessing region (as
would be suggested by the small inferred $\Omega=0.05$) and thus
limits the blurring of the reprocessed profile.  Alternatively, if we
are viewing the hard beam from the neutron star off-axis, but the beam
strikes the disk directly, then the reprocessing region may see
pulsations with a significantly higher amplitude than what we observe.

We conclude that the interpretations regarding the disk and beam
geometries are very different depending on whether we consider the
$E>10$ keV X-rays which comprise the bulk of the surface emission from
the neutron star.  Further, we note that the complex pulse profile and
lower pulse fraction in the medium band relative to the hard band
suggest that the neutron star emission not be described by a simple
pencil beam at all energies.  Complex beam shapes have been proposed
for some X-ray pulsars; for example, the hard pulses from Her X-1 have
been modeled as a central pencil beam surrounded by a fan beam
\citep{blum00}.  In our \suzaku\ observations of LMC X-4, the only
clear way to explain changes in the medium energy pulse profile (particularly
the change from single to double peaks, and the relatively weak
pulsations in observation 3) is by intrinsic changes in the beam
pattern that move the beam out of our line of sight.  Weak or absent
pulsations have been observed before in the high state \citep{laba01},
so perhaps the changes in the beam pattern are fairly common and
happen on timescales of $\sim$ weeks.

Despite the difficulty in interpreting the observed pulse profiles,
these observations provide further evidence for the soft component
originating in a reprocessing disk.  As mentioned in
Section~\ref{subsec:spectroscopy}, using $R_{\rm BB}\sim 1.5\times
10^{8}\ \rm cm$ and $B \sim 10^{13}$ G, we obtain $R_{\rm BB}/R_{m} =
0.7$.  Keeping in mind that many uncertainties exist in calculating
$R_{\rm BB}$ and $R_{m}$ \citep[such as difficulties in measuring $B$
  and the poorly understood interaction of the accretion disk with the
  magnetic field; see, e.g.][]{scot00, hick04}, the observed
correspondence between $R_{\rm BB}$ and $R_{\rm m}$ is rather striking
and suggests that the hard X-rays are reprocessed by optically thick
material near the magnetosphere.

Finally, we reiterate that the interpretation of disk reprocessing for
the soft component is preferred on physical grounds.  As discussed in
Section~\ref{subsec:spectroscopy}, a thermal bremsstrahlung or MEKAL model (rather
than a blackbody) adequately describes the soft spectral component in
LMC X-4, but we can rule out this mechanism because it would require
too large of an emission region.  We conclude, as \citet{hick05} have
inferred for SMC X-1, the soft X-ray component of LMC X-4 is analogous
to that of Her X-1 and hence due to reprocessing of hard X-rays by the
inner accretion disk.

As an aside, we note that if for LMC X-4 we consider the ``hard'' flux
only at 2--10 keV, we obtain a smaller $L_X$ and thus for $R_{\rm
  BB}/R_{m}\sim 0.3$, which is in the range of the values  ($R_{\rm BB}/R_{m}\sim 0.2$--0.4) measured for
Her X-1 over
similar energy bands \citep{scot00,hick04}.  This suggests that $R_{\rm BB}$ may be somewhat
closer to $R_{\rm m}$ for Her X-1 as well, if we consider the full
luminosity of the hard component which peaks at higher X-ray energies.

A more detailed understanding of the emission from LMC X-4 will
require pulse-phase-resolved spectroscopy throughout the superorbital
period, as well as a more sophisticated model of the beam and disk
geometry. In the future, the {\em International X-ray
  Observatory}\footnotemark, with its greater spectral resolution and
effective area, may enable us to undertake detailed, high-resolution
pulse-phase spectroscopy and extend our studies to the low state of
the superorbital cycle, enabling us to probe further into the
accretion mechanism of LMC X-4, and gain a better understanding of
accretion in highly magnetized environments.
\footnotetext{http://ixo.gsfc.nasa.gov}


\acknowledgements 

We are grateful to Joseph Neilsen for his contribution to the data
analysis, and Chris Done, Christine Jones, Jonathan McDowell, and
Marie Machacek for helpful discussions.  We also thank the Suzaku
Helpdesk for their valuable advice on the data analysis.  The work was
supported in part by the National Science Foundation Research
Experiences for Undergraduates (REU) and Department of Defense Awards
to Stimulate and Support Undergraduate Research Experiences (ASSURE)
programs under grant 0754568, Suzaku grant NNX08AI17G, ADP grant
NNX08AJ61G, and by the Smithsonian Institution.  R.C.H. was supported by
an SAO Postdoctoral Fellowship and an STFC Postdoctoral Fellowship.


\end{document}